\documentclass[twocolumn]{aa}
\usepackage[pdftex]{graphics}
\usepackage{graphicx}
\usepackage{txfonts}
\usepackage{ifpdf}
\usepackage[round]{natbib}
\newcommand{\myarcsec}{\hbox{$.\!\!^{\prime\prime}$}}
\newcommand{\myarcmin}{\hbox{$.\!\!^{\prime}$}}
\newcommand{\bolda}{a}
\newcommand{\bolde}{e}
\newcommand{\boldg}{g}
\newcommand{\be}{\begin{equation}}
\newcommand{\ee}{\end{equation}}
\newcommand{\dd}{{\rm{d}}}

\def\deg{\hbox{$^\circ$}}
\def\sun{\hbox{$\odot$}}

\begin{document}

   \title{ARCRAIDER I: Detailed optical and X-ray analysis of the cooling flow cluster Z3146\thanks{Based on observations made with the NASA/ESA Hubble Space Telescope, obtained
   from the data archive at the Space Telescope Institute (PID-number 8301). STScI is operated by the
   association of Universities for Research in Astronomy, Inc. under
   the NASA contract NAS 5-26555. Also based on observations made with ESO Telescopes
   at the La Silla or Paranal Observatories under programme ID 68.A-02555 and 073.A-0050 and on
   observations with XMM-Newton, an ESA Science Mission with instruments and contributions
   directly funded by ESA Member states and the USA (NASA).}}

   \author{W. Kausch\inst{1}
      \and
      M. Gitti\inst{2}
      \and
      T. Erben\inst{3}
      \and
      S. Schindler\inst{1}
%      \and
%      A. Schwope\inst{3}
%      \and
%      J. Wambsganss\inst{4}
          }

   \offprints{W. Kausch}

   \institute{Institut f\"ur Astro- und Teilchenphysik, University of Innsbruck,
             Technikerstr. 25, A-6020 Innsbruck, Austria\\
             \email{wolfgang.kausch@uibk.ac.at}
  %       \and
  %          Dept. of Physics \& Astronomy, Ohio University, Clippinger Lab, Athens, OH
  %          45701, USA
         \and
            INAF - Osservatorio Astronomico di Bologna, via Ranzani 1, 40127 Bologna,
            Italy
         \and
             Argelander-Institut f\"ur Astronomie (AIfA), University of Bonn,
         Auf dem H\"ugel 71, D-53121 Bonn, Germany
             }

   \date{Received; accepted}

   \abstract{
We present a detailed analysis of the medium redshift ($z=0.2906$)
galaxy cluster Z3146 which is part of the ongoing ARCRAIDER project,
a systematic search for gravitational arcs in massive clusters of
galaxies. The analysis of Z3146 is based on deep optical wide field
observations in the $B$, $V$ and $R$ bands obtained with the
WFI@ESO2.2m, and shallow archival WFPC2@HST taken with the $F606W$
filter, which are used for strong as well as weak lensing analyses.
Additionally we have used publicly available XMM/Newton observations
for a detailed X-ray analysis of Z3146. Both methods, lensing and
X-ray, were used to determine the dynamical state and to estimate
the total mass. We also identified four gravitational arc candidates.\\
We find this cluster to be in a relaxed state, which is confirmed by
a large cooling flow with nominal $\sim 1600$\,M$_\odot$ per year,
regular galaxy density and light distributions and a regular shape
of the weak lensing mass reconstruction. The mass content derived
with the different methods agrees well within 25\% at
$r_{200}=1661\;h_{70}^{-1}$kpc indicating a velocity dispersion of
$\sigma_v=869^{+124}_{-153}$\,km/s.

   }

   \maketitle
%
%_______________________________________________________________________________________________________________________________________
\section{Introduction}
Galaxy clusters are the largest bound structures in the universe and
therefore excellent cosmological probes. In particular, large
samples of clusters allow a statistical study of their physical
properties. Such samples need clear selection criteria, e.g.
selection by mass. Due to the tight relation between the X-ray
luminosity $L_{\rm X}$ and the mass \citep{schindler3,reiprich}
X-ray surveys provide an excellent basis to select the most massive
systems for lensing studies
\citep{luppino99a,smith01a,smith05a,bardeau05a}.\\
The combination of lensing and X-ray studies allows us to get
important insights into galaxy clusters, \rm as it offers the
possibility to obtain physical properties of the cluster members,
the Intra Cluster Medium (ICM) and the determination of the cluster
gravitational mass and its distribution with independent methods.\\
However, the mass determination of galaxy clusters is a very
difficult task. It is dependent on the method adopted and on the
validity of the assumptions used to convert observables to cluster
masses. Currently two methods are widely used: (a) From the gas
density and temperature profiles measured with X-ray observations it
is possible to derive an estimate of the gravitational mass by
assuming spherical symmetry and hydrodynamical equilibrium
\citep[see
e.g.][]{allen01a,allen02a,ettori03a,pointecouteau04a,pratt05a,voigt06a}.
(b) The second method is based on gravitational lensing analyses,
using either strongly deformed background sources (arcs) to
constrain the cluster mass in the very cluster centre or statistical
methods to investigate systematic shape distortions of background
objects to map the mass distribution of a cluster (weak lensing
method, see e.g. \cite{weaklensing} for a review on this topic). The
lensing method is affected by the least number of assumptions as it
is neither sensitive to the nature of the matter nor its dynamical
state. However, this method measures the integrated mass along the
line of sight, which can lead to a bias of too high mass estimates
\citep{white02a}. Detailed lensing analyses were carried out for
several galaxy clusters, e.g. CL0024+1654
\citep{kneib03a,czoske02a}, A2218 \citep{kneib96a},
A1689 \citep{broadhurst05a,broadhurst05b}, A383 \citep{smith01a} or RX J1347-1145 \citep{bradac05a}.\\
Unfortunately the mass estimates derived from the different methods
can be quite inconsistent. In some clusters there are considerable
discrepancies up to a factor of 3, e.g. MS0440+0204 \citep{gioia98a}
or CL0500-24 \citep{schindler3}. \cite{allen98a} found the lensing
and X-ray method to be consistent for cooling flow clusters, whereas
for non-cooling flow clusters the mass discrepancy between the
strong lensing method and the X-ray based mass determinations can
differ by a factor of up to $2-4$. This mainly comes from the fact
that the inner core of clusters, where strong lensing occurs, is not
well described by the usual simple models used in X-ray methods,
which are based on the assumptions mentioned above. The
discrepancies of the weak lensing and the X-ray method seem to be
much smaller \citep{wu98a}.

In this paper we present a combined optical, X-ray and lensing
analysis of Z3146. This cluster of galaxies is located at
$\alpha=$10h23m39.6s, $\delta=+04^\circ11\arcmin10\arcsec$ (J2000)
with a redshift of $z_{\rm cl}=0.2906$ \citep{schwope} and was the
subject of many previous optical \citep[e.g.
][]{crawford99a,edge02a,chapman02a,edge03a,sand05a} and X-ray
investigations \citep[e.g. ][]{edge94a,
ettori01a,fabian02a,hicks05a}. This prominent cluster is one of the
most X-ray luminous systems in the ROSAT Bright Survey
\citep[hereafter RBS]{schwope} having an X-ray luminosity of about
$\log(L_{\rm X})=45.3$\,erg/s in the 0.5-2\,keV ROSAT band. It is
part of a larger sample of X-ray selected galaxy clusters which is
described in Sect. \ref{sec.ARC_RAIDER} and will be given in more
detail in a forthcoming paper (Kausch et al., in prep.). This paper
contains detailed X-ray and lensing analyses of Z3146 followed by
several optical investigations. The X-ray analysis is presented in
Sect. \ref{sec:xray}, a description of the optical observations and
the data reduction procedure used for this investigation is given in
Sect. \ref{sec:opt_observations}. Sect. \ref{sec:opt_analysis}
contains a lensing analysis based on weak (Sect. \ref{subsect:wl})
and strong lensing (Sect. \ref{sec:stronglensing}).
Sect.\,\ref{sec:furtherinvestigations} comprises continuative
optical investigations on the cluster. In section Sect. \ref{sec:summary}
we summarize and discuss the results.\\
Throughout this paper we use H$_0=70\,h_{70}$\,kms$^{-1}$Mpc$^{-1}$
and $\Omega_M=1-\Omega_\Lambda=0.3$. Hence
$1\myarcsec0\hat{=}4.36\,\,h_{70}^{-1}$\,kpc for the cluster
redshift of $z=0.2906$.

%_______________________________________________________________________________________________________________________________________
\section{Description of the ARCRAIDER-Project}
\label{sec.ARC_RAIDER} Z3146 is part of the ARCRAIDER sample of
galaxy clusters (Kausch et al. in prep.). ARCRAIDER stands for \bf
ARC\rm statistics with X-\bf RA\rm y lum\bf I\rm nous me\bf D\rm ium
r\bf E\rm dhift galaxy cluste\bf R\rm s. The project is based on a
homogeneous and unique sample of galaxy clusters chosen from the RBS
\citep{schwope}, a compilation of all X-ray sources with a PSPC
count rate $>0.2$\,s$^{-1}$. As all sources are located at high
galactic latitudes ($|b|>30\deg$), the $n_{\rm H}$ values for our
clusters are very small ($n_{\rm H}\leq7.7\times10^{20}$cm$^{-2}$).
The selected clusters satisfy the following criteria: (a) located in
the medium redshift range $0.1\leq z\leq0.52$, (b) an X-ray
luminosity $\geq0.5\times10^{45}$\,erg/sec (0.5-2\,keV band), (c)
classified as clusters in the ROSAT Bright Survey, (d) not a member
of the Abell catalogue, and (e) visible from La Silla/Paranal
(declination $\delta\leq20^\circ$).\\
The total sample contains 22 galaxy clusters which were observed
with different telescopes: RBS1316 (RX J1347-1145) is the most X-ray
luminous cluster known \citep[][Gitti, Piffaretti \& Schindler 2007,
in prep.]{schindler2,allen02a,gitti04a} and was observed in the $U$,
$B$, $V$, $R$ and $I$ band with the ESOVLT with the FORS1 instrument
and in the $K_s$ band with ISAAC \citep{bradac05a}. All other
clusters were observed at least in the $V$ and $R$ band either with
the SUperb Seeing Imager 2 (SUSI2@ESONTT, ESO-filters V\#812 and
R\#813) or with the Wide Field Imager (BB\#V/89\_ESO843 and
WFI@ESO2.2m, ESO-filters BB\#R$_{\rm c}$/162\_ESO844) with usually
half the exposure time in $V$ than in $R$. We use the deep $R$ band
frame as the primary science band for our lensing analysis, whereas
the shallow $V$ image is used for
colour determinations for a rough division between foreground and background galaxies.\\
As our clusters are the most luminous ones of the RBS, we expect
these systems to be very massive due to the $L_{\rm X}-M$ relation
\citep{reiprich,schindler3}. Therefore it is very likely to find
strong gravitational lensing features like arcs or arclets in such
systems. A similar sample of clusters was established by
\cite{luppino99a}, based on the EMSS. In total they found arc(lets)
and candidates in $\sim42$\% of their members. As their X-ray
luminosity limit was chosen to be lower than ours ($L_{\rm
X}>2\cdot10^{44}$\,erg/sec in the 0.3-3.5\,keV regime) we expect
to detect gravitational arcs \rm in 45-60\% of the clusters.\\

%_______________________________________________________________________________________________________________________________________
\section{X--ray analysis of Z3146} \label{sec:xray}
\subsection{Observation and data preparation}

Z3146 was observed by \textit{XMM--Newton} in December 2000 during
rev. 182 (PI: Mushotzky) with the MOS and pn detectors in Full Frame
Mode with THIN filter, for an exposure time of 53.1 ks for MOS and
46.1 ks for pn. We used the SASv6.0.0 processing tasks
\textit{emchain} and \textit{epchain} to generate calibrated event
files from raw data. Throughout this analysis single pixel events
for the pn data (PATTERN 0) were selected, while for the MOS data
sets the PATTERNs 0-12 were used. The removal of bright pixels and
hot columns was done in a conservative way applying the expression
(FLAG==0). To reject the soft proton flares we accumulated the light
curve in the [10-12] keV band for MOS and [12-14] keV band for pn,
where the emission is dominated by the particle--induced background,
and excluded all the intervals of exposure time having a count rate
higher than a certain threshold value (the chosen threshold values
are 15 counts/100 s for MOS and 20 counts/100 s for pn). The
remaining exposure times after cleaning are 52.3 ks for MOS1, 52.6
ks for MOS2 and 45.7 ks for pn. Starting from the output of the SAS
detection source task, we made a visual selection on a wide energy
band MOS \& pn image of point sources in the FoV. Events from these
regions were excluded directly from each event list.

The background estimates were obtained using a blank-sky observation
consisting of several high-latitude pointings with sources removed
\citep{lumb02a}. The blank-sky background events were selected using
the same selection criteria (such as PATTERN, FLAG, etc.), intensity
filter (for flare rejection) and point source removal used for the
observation events; this yields final exposure times for the blank
fields of 365 ks for MOS1, 350 ks for MOS2 and 294 ks for pn. Since
the cosmic ray induced background might slightly change with time,
we computed the ratio of the total count rates in the high energy
band ([10-12] keV for MOS and [12-14] keV for pn). The obtained
normalization factors (0.827, 0.820, 0.836 for MOS1, MOS2 and pn,
respectively) were then used to renormalize the blank field data.
The blank-sky background files were recast in order to have the same
sky coordinates as Z3146. For the pn data, we generated a list of
out-of-time events (hereafter OoT) to be treated as an additional
background component. The effect of OoT in the current observing
mode (Full Frame) is 6.3\%. The OoT event list was processed in a
similar way as done for the pn observation event file. The
background subtraction (for spectra and surface brightness profiles)
was performed as described in
\cite{arnaud02a}. In case of pn the OoT data were also subtracted.\\
The source and background events were corrected for vignetting using
the weighted method described in \cite{arnaud01a}, the weight
coefficients being tabulated in the event list with the SAS task
\textit{evigweight}. This allows us to use the on-axis response
matrices and effective areas.
\\
Unless otherwise stated, the reported errors are at 90\% confidence
level in the entire Sect.\,\ref{sec:xray}.

%%%%%%%%%%%%%%%%%%%%%%%%%%%%%%%%%%%%%%%%%%%%%%%%%%%%%%%%%%%%%%%%%%%%%%%%%%%%%%

\subsection{Morphological analysis}
\label{morphology.sec}

\begin{figure}[ht]
\includegraphics[angle=0,width=8cm]{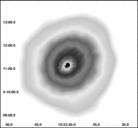}
%\special{psfile=figure/zw3146-bw.eps hoffset=0 voffset=-255
%vscale=40 hscale=40 angle=0} \vspace{7.5cm}
\caption{ MOS1 image of
ZW3146 in the [0.3-10] keV energy band. The image is corrected for
vignetting and exposure and is adaptively smoothed (signal-to-noise
ratio = 20). } \label{image}
\end{figure}

The adaptively smoothed, exposure corrected MOS1 count rate image in
the [0.3-10] keV energy band is presented in Fig. \ref{image}. The
smoothed image was obtained from the raw image corrected for the
exposure map (that accounts for spatial quantum efficiency, mirror
vignetting and field of view) by running the task \textit{asmooth}
set to a desired signal-to-noise ratio of 20. Regions exposed with
less than 10\% of the total exposure were not considered.

We notice a sharp central surface brightness peak at a position
$10^{\rm h} 23^{\rm m} 40^{\rm s}.01 +04^{\circ} 11' 09\myarcsec45$
(J2000), in very good agreement ($\Delta\alpha=0\myarcsec41$,
$\Delta\delta=0\myarcsec55$) with the optical position of the
central dominant cluster galaxy \citep{schwope}. The morphology of
the cluster is quite regular, thus indicating a relaxed dynamical
state, even though we notice that the central core appears
slightly shifted %$\sim$ 10 arcsec
to the south-east with respect to the outer envelope, with a
north-west to south-east elongation of the cluster core. The regular
morphology of the cluster is indicative of a relaxed dynamical
state, thus allowing us to derive a good mass estimate based on the
usual assumptions of hydrostatic equilibrium and spherical symmetry
(see Sect. \ref{mass.sec}).

\subsubsection{Surface brightness profile}
\label{brightness.sec}

\begin{figure}[ht]
%\special{psfile=lowres2/count_rate.eps hoffset=0 voffset=-240
%vscale=80 hscale=80 angle=0} \vspace{8.5cm}
\includegraphics[angle=0,width=8cm]{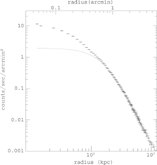}
\caption{ Background subtracted, azimuthally averaged radial surface
brightness profile in the [0.3-2] keV range. The best fit
$\beta$-model fitted over the 230 - 1300 kpc region
(0\myarcmin9-5\myarcmin0) is over-plotted as a solid line. When
extrapolated to the centre, this model shows a strong deficit as
compared to the observed surface brightness. } \label{1betafit}
\end{figure}

We computed a background-subtracted vignetting-corrected radial
surface brightness profile in the [0.3-2] keV energy band for each
camera separately. The profiles for the three detectors were then
added into a single profile and binned such that at least a
sigma-to-noise ratio of 3 was reached. The cluster emission is
detected up to 1.5 Mpc ($\sim 6\arcmin$) and the profile appears
relatively regular and relaxed (see Fig. \ref{1betafit}).
%
%\begin{figure}[ht]
%\special{psfile=figure/fit-2beta.ps hoffset=0 voffset=-260 vscale=40
%hscale=40 angle=0}
%\vspace{7.0cm}
%\caption{
%Background subtracted, azimuthally averaged radial surface brightness profile
%in the [0.3-2] keV range. The best fit double isothermal $\beta$-model
%fitted over the 15 - 1300 kpc region is over-plotted as a solid line.}
%\label{2betafit}
%\end{figure}
%
The surface brightness profile of the undisturbed cluster was fitted
with the CIAO tool \textit{Sherpa} with various parametric models,
which were convolved with the \textit{XMM-Newton} PSF. The overall
PSF was obtained by adding the PSF of each camera
\citep{ghizzardi01}, estimated at an energy of 1.5 keV and weighted
by the respective cluster count rate in the 0.3-2 keV energy band. A
single $\beta$-model \citep{cavaliere76a} is not a good description
of the entire profile: a fit to the outer regions
(230$\;h_{70}^{-1}$\,kpc $< r <$ 1300$\;h_{70}^{-1}$\,kpc) shows a
strong excess in the centre as compared to the model (see Fig.
\ref{1betafit}). The peaked emission is a strong indication for a
cooling flow in this cluster. We found that for
230$\;h_{70}^{-1}$\,kpc $< r <$ 1300$\;h_{70}^{-1}$\,kpc the data
can be described ($\chi^2_{\rm red} \sim 1.39$ for 74 d.o.f.) by a
$\beta$-model with a core radius $r_{\rm c}=177 \pm
2$$\;h_{70}^{-1}$\,kpc and a slope parameter $\beta=0.77 \pm 0.01$
(3 $\sigma$ confidence level). The single $\beta$-model functional
form is a convenient representation of the gas density profile in
the outer regions, which is used as a tracer for the potential. The
parameters of this best fit are thus used in the following to
estimate the total mass profile in the region where the single beta
model holds (see Sect. \ref{mass.sec}).\\
We also considered a double isothermal $\beta$-model and found that
it can account for the entire profile, if the very inner and outer
regions are excluded:
%(see Fig. \ref{2betafit}):
for 15 kpc$\;h_{70}^{-1}$\,$< r <$ 1300$\;h_{70}^{-1}$\,kpc the best
fit parameters are $r_{\rm c1}=177 \pm 2$$\;h_{70}^{-1}$\,kpc,
$r_{\rm c2}=39 \pm 1$$\;h_{70}^{-1}$\,kpc and $\beta=0.76 \pm 0.01$
($\chi^2_{\rm red} \sim 1.67$ for 95 d.o.f.). A common $\beta$ value
is assumed in this model, but we also tried the fit with two
different  $\beta$ values, finding very similar results ($r_{\rm
c1}=172 \pm 2$$\;h_{70}^{-1}$\,kpc, $r_{\rm c2}=45 \pm
1$$\;h_{70}^{-1}$\,kpc, $\beta_1=0.76 \pm 0.01$ and $\beta_2=0.88
\pm 0.01$; $\chi^2_{\rm red} \sim 1.68$ for 94 d.o.f.).

%%%%%%%%%%%%%%%%%%%%%%%%%%%%%%%%%%%%%%%%%%%%%%%%%%%%%%%%%%%%%%%%%%%%%%%%%%%%%%%

\subsection{Temperature map}
\label{tmap.sec}

\begin{figure}[ht]
%\special{psfile=lowres2/zw3146-Tmap.ps hoffset=-10 voffset=-290
%vscale=45 hscale=45 angle=0} \vspace{7.5cm}
\includegraphics[angle=0,width=8cm]{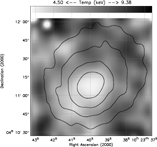}
\caption{ Temperature map obtained by using 4 X-ray colours
(0.3-1.0, 1.0-2.0, 2.0-4.5, 4.5-8 keV) and estimating the expected
count rate with XSPEC for a thermal MEKAL model, with fixed Galactic
absorption $N_H$ and metallicity $Z$. } \label{tmap.fig}
\end{figure}

The temperature image of the cluster central region shown in Fig.
\ref{tmap.fig} was built from X-ray colours. Specifically, we
extracted mosaiced MOS images in four different energy bands
(0.3-1.0 keV, 1.0-2.0 keV, 2.0-4.5 keV and 4.5-8 keV), subtracted
the background and divided the resulting images by the exposure
maps. A temperature in each pixel of the map was obtained by fitting
values in each pixel of these images with a thermal plasma, fixing
$n_{\rm H}$ to the Galactic value and the element abundance to 0.3
solar. In particular we note that the very central region is cooler
than the surrounding medium and the north-west quadrant appears
slightly hotter than the south-east one, even though no strong
features are present.
The regularity of the temperature distribution points to a relaxed
dynamical state of the cluster, thus excluding the presence of an
ongoing merger. Since cluster merging can cause strong deviations
from the assumption of an equilibrium configuration, this allows us
to derive a good estimate of the cluster mass (see Sect.
\ref{mass.sec}).

%%%%%%%%%%%%%%%%%%%%%%%%%%%%%%%%%%%%%%%%%%%%%%%%%%%%%%%%%%%%%%%%%%%%%%%%%%%%%%%

\subsection{Spectral analysis}
\label{spectral.sec}

Throughout the analysis, a single spectrum was extracted for each
region of interest and was then regrouped to reach a significance
level of at least 3 $\sigma$ in each bin. The data were modelled
using the XSPEC code, version 11.3.0. Unless otherwise stated, the
relative normalizations of the MOS and pn spectra were left free
when fitted simultaneously. We used the following response matrices:
{\ttfamily m1{\_}169{\_}im{\_}pall{\_}v1.2.rmf} (MOS1), {\ttfamily
m2{\_}169{\_}im{\_}pall{\_}v1.2.rmf} (MOS2), {\ttfamily
epn{\_}ff20{\_}sY9.rmf} (pn).

\subsubsection{Global spectrum analysis}
\label{global.sec}

\begin{figure}[ht]
\centerline{\includegraphics[angle=0,width=\hsize]{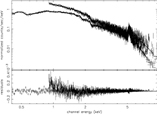}}
\caption{ Global MOS (lower) and pn (upper) spectra in the
[0.4/0.9-10.0] keV energy range (MOS/pn) integrated in a circular
region of radius 5$\arcmin$. The fit with a MEKAL model and the
residuals are shown. } \label{spettro}
\end{figure}

For each instrument, a global spectrum was extracted from all events
lying within 5$\arcmin$ of the cluster emission peak. We tested in
detail the consistency between the three cameras by fitting
separately these spectra with an absorbed MEKAL model with the
redshift fixed at z=0.291 and the absorbing fixed at the galactic
value \citep[$n_{\rm H} = 3.01 \times 10^{20} \mbox{ cm}^{-2}$,
][]{dickey90a}. Fitting the data from all instruments above 0.3 keV
led to inconsistent values for the temperature derived with the MOS
and pn cameras: $kT = 6.18^{+0.16}_{-0.15}$ keV (MOS1),
$5.72^{+0.15}_{-0.14}$ (MOS2), $5.01^{+0.09}_{-0.08}$ (pn). We then
performed a systematic study of the effect of imposing various high
and low-energy cutoffs, for each instrument. Good agreement between
the three cameras was found in the [1.0-10.0] keV energy range ($kT
= 6.27^{+0.21}_{-0.20}$ keV for MOS1, $5.99^{+0.20}_{-0.19}$ for
MOS2, $6.00^{+0.16}_{-0.15}$ for pn). On the other hand, we also
found consistent results by fitting the MOS spectra in the [0.4-10]
keV energy range and the pn spectrum in the [0.9-10] keV energy
range.The discrepancies observed by fitting the whole energy range
are probably due to some residual calibration uncertainties in the
low-energy response of all instruments. Thus, in order to avoid
inaccurate measurements due to calibration problems, we adopted the
low energy cut-off derived above for the
spectral analysis discussed below.\\
The combined MOS+pn global temperature and luminosity are
respectively $kT = 5.91^{+0.09}_{-0.08}$ keV, $L_X \mbox{ (2-10
keV)}= 2.0 \pm 0.1 \times 10^{45} \mbox{ erg s}^{-1}$ in the
[0.4/0.9-10.0] keV energy range (MOS/pn) and $kT =
6.08^{+0.11}_{-0.10}$ keV, $L_X \mbox{ (2-10 keV)}= 2.0 \pm 0.1
\times 10^{45} \mbox{ erg s}^{-1}$ in the [1.0-10.0] keV energy
range. These values are in agreement with ASCA results (Allen et al.
1996, Allen 2000), while Ettori et al. (2001) derived higher
temperature values from \textit{BeppoSAX} observations. The
simultaneous fit in the [0.4/0.9-10.0] keV energy range (MOS/pn) to
the three spectra is shown in Fig. \ref{spettro}.

%\begin{figure}[ht]
%\special{psfile=figure/spettro.ps hoffset=-20 voffset=0 vscale=33
%hscale=33 angle=-90} \vspace{7.cm} \caption{ Global MOS (lower) and
%pn (upper) spectra in the [0.4/0.9-10.0] keV energy range (MOS/pn)
%integrated in a circular region of radius 5$'$. The fit with a MEKAL
%model and the residuals are shown. } \label{spettro}
%\end{figure}

\subsubsection{Radial temperature profile}
\label{radial.sec}

\begin{table*}
\begin{center}
\hskip 0.2truein \caption { Results from the spectral fitting in
concentric annular regions in the [0.4/0.9-10.0] keV energy range
(MOS/pn) and in the [1.0-10.0] keV energy range. Temperatures ($kT$)
are in keV, metallicities ($Z$) in solar units and [2-10] keV
luminosities ($L_{\rm X}$) in units of $10^{44} \mbox{ erg s}^{-1}$.
The total $\chi^2$ values and numbers of degrees of freedom (dof) in
the fits are also listed. Error bars are at the 90\% confidence
levels on a single parameter of interest. } \vskip 0.1truein \hskip
0.0truein
\begin{tabular}{cc|cccc|cccc}
~ & ~ & \multicolumn{4}{|c|}{[0.4/0.9-10.0] keV energy range (MOS/pn)} & \multicolumn{4}{|c}{[1.0-10.0] keV energy range} \\
~&~&~&~&~&~&~&~&~\\
Radius & Radius (kpc)   &     $kT$    & $Z$     &  $L_{\rm X}$ &     $\chi^2$/dof  & $kT$    & $Z$    &   $L_{\rm X}$ &     $\chi^2$/dof   \\
\hline
~&~&~&~&~&~&~&~&~&~\\
0-20$''$ & 0-87 & $4.7^{+0.1}_{-0.1}$ & $0.38^{+0.03}_{-0.03}$ & 5.68 & 1630/1503 & $4.8^{+0.1}_{-0.1}$ & $0.36^{+0.04}_{-0.03}$ & 5.77  & 1461/1403 \\
20$''$-40$''$ & 87-174 & $6.1^{+0.2}_{-0.2}$ & $0.31^{+0.04}_{-0.04}$ & 5.15 & 1456/1421 & $6.3^{+0.2}_{-0.2}$ & $0.30^{+0.04}_{-0.04}$ & 5.20 & 1323/1321 \\
40$''$-1$'$ & 174-262 & $6.7^{+0.3}_{-0.3}$ & $0.25^{+0.05}_{-0.05}$ & 3.14 & 1145/1139 & $ 6.9^{+0.3}_{-0.3}$ & $0.25^{+0.05}_{-0.05}$ & 3.16 & 1040/1039 \\
1$'$-1\myarcmin5 &262-393 & $6.5^{+0.3}_{-0.3}$ & $0.18^{+0.05}_{-0.05}$ & 2.41 & 1002/1007 & $ 6.7^{+0.4}_{-0.3}$ & $0.18^{+0.05}_{-0.05}$ & 2.43 & 892/907 \\
1\myarcmin5-2$'$ &393-524 & $6.7^{+0.5}_{-0.4}$ & $0.28^{+0.08}_{-0.08}$ & 1.28 & 743/719 & $ 7.0^{+0.6}_{-0.5}$ & $0.28^{+0.08}_{-0.08}$ & 1.30 & 618/620 \\
2$'$-3$'$ &524-785 & $7.0^{+0.6}_{-0.6}$ & $0.24^{+0.10}_{-0.10}$ & 1.21 & 593/578 & $ 7.4^{+0.8}_{-0.7}$ & $0.22^{+0.10}_{-0.10}$ & 1.26 & 479/480 \\
3$'$-5$'$ &785-1309 & $6.8 ^{+1.3}_{-0.9}$ & $0.23^{+0.28}_{-0.22}$ & 0.79 & 306/253 & $ 7.1^{+1.9}_{-1.2}$ & $0.21^{+0.30}_{-0.21}$ & 0.80 & 241/195 \\
0$'$-5$'$ &0-1309 & $5.9^{+0.1}_{-0.1}$ & $0.30^{+0.02}_{-0.02}$ & 20.14 & 2632/1859 & $6.1^{+0.1}_{-0.1}$ & $0.28^{+0.02}_{-0.02}$ & 20.42 & 2399/1759\\
\hline
\end{tabular}
\label{profilot.tab}
\end{center}
\end{table*}

\begin{figure}[ht]
%\special{psfile=figure/profiloT.ps hoffset=0 voffset=-260 vscale=40
%\special{psfile=lowres2/temperature.ps hoffset=0 voffset=0
%vscale=100 hscale=100 angle=0} \vspace{8.5cm}
\includegraphics[angle=0,width=\hsize]{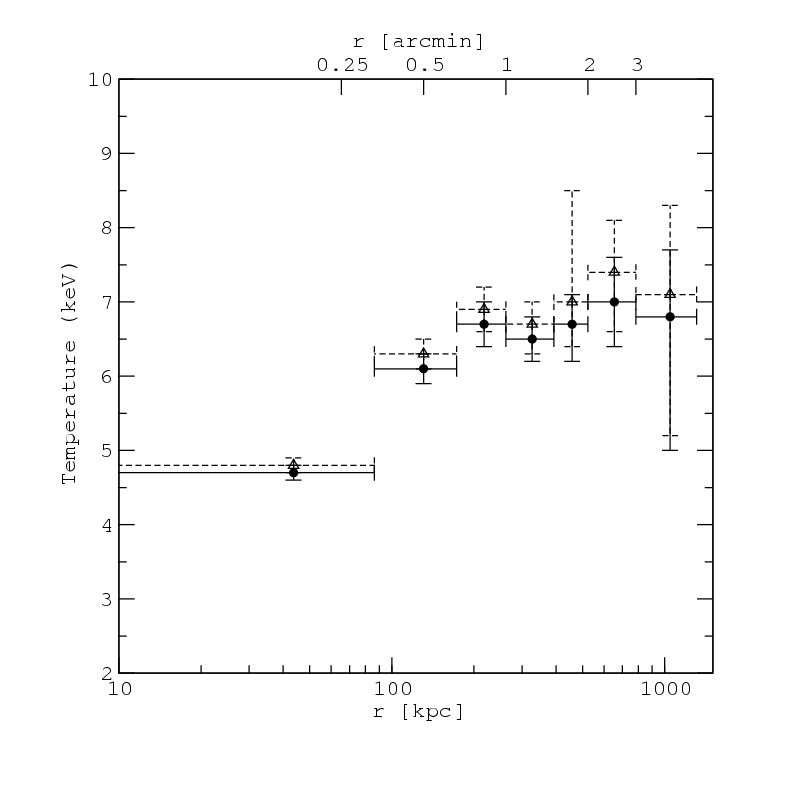}
\caption{ \textit{Full Circles}: the projected X-ray gas temperature
profile measured from \textit{XMM} data in the [0.4/0.9-10.0] keV
energy range (MOS/pn). \textit{Open Triangles}: same as full
circles, but in the [1.0-10.0] keV energy range. }
\label{profilot.fig}
\end{figure}

\begin{figure}[ht]
%\includegraphics[angle=0,width=9cm]{figure/met-profilo.ps}
%\special{psfile=lowres/metallicity.ps hoffset=0 voffset=-260
%vscale=35 hscale=35 angle=0} \vspace{8.5cm}
\includegraphics[angle=0,width=8cm]{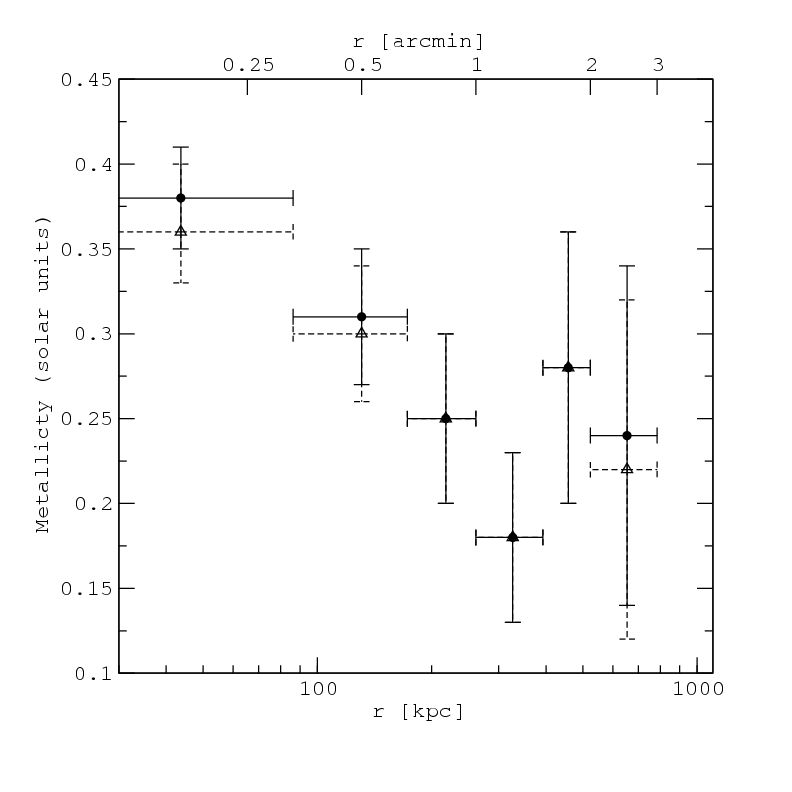}
\caption{ \textit{Full Circles}: the projected X-ray gas metallicity
profile measured from \textit{XMM} data in the [0.4/0.9-10.0] keV
energy range (MOS/pn). \textit{Open Triangles}: same as full
circles, but in the [1.0-10.0] keV energy range. }
\label{profilomet.fig}
\end{figure}

We produced a radial temperature profile by extracting spectra in
annuli centred on the peak of the X--ray emission. The annular
regions are detailed in Table\,\ref{profilot.tab}. The data from the
three cameras have been modelled simultaneously using a simple,
single-temperature model (MEKAL plasma emission code in XSPEC) with
the absorbing column density fixed at the nominal Galactic value.
%($n_{\rm H}=3.01 \times 10^{20} \mbox{ atom cm}^{-2}$, Dickey\&Lockman 1990).
The free parameters in this model are the temperature $kT$,
metallicity $Z$ (measured relative to the solar values, with the
various elements assumed to be present in their solar ratios) and
normalization (emission measure). We separately performed the
spectral fitting in the [0.4/0.9-10.0] keV energy range (MOS/pn) and
in the [1.0-10.0] keV energy range. The best-fitting parameter
values and 90\% confidence levels derived from the fits to the
annular spectra are summarized in Table\,\ref{profilot.tab}. The
projected temperature profile determined with this model is shown in
Fig. \ref{profilot.fig}. We note that, as expected, temperature
values derived in the [1.0-10.0] keV energy range are slightly
higher than those derived in the [0.4/0.9-10.0] keV energy range
(MOS/pn), even though consistent within the 90\% confidence level.
In the following discussion we adopt results derived in the
[0.4/0.9-10.0] keV energy range (MOS/pn). The temperature rises from
a mean value of $4.7 \pm 0.1$ keV within 90$\;h_{70}^{-1}$\,kpc to
$kT = 6.7 \pm 0.5$ keV over the 180-1300 kpc region, where the
cluster can be considered approximately isothermal. The lack of
evidence for a temperature decline in the outer regions is in
agreement with the results by Mushotzky (2003).
\\
~\\
The metallicity profile is shown in Fig. \ref{profilomet.fig}: a
gradient is visible towards the central region, the metallicity
increasing from $Z=0.18 \pm 0.05$  over the
260-400$\;h_{70}^{-1}$\,kpc region to $Z=0.38 \pm 0.03$ inside the
central 90$\;h_{70}^{-1}$\,kpc.

\subsubsection{Cooling core analysis}
\label{cf.sec}

The surface brightness profile, the temperature map and the
temperature profile all give hints of the presence of a cooling
core. Here we further investigate the physical properties of the ICM
in the central region. The cooling time is calculated as the time
taken for the gas to radiate its enthalpy per unit volume using the
instantaneous cooling rate at any temperature:
\begin{equation}
t_{\rm cool} \approx \frac{H}{n_{\rm e} n_{\rm H} \Lambda(T)} =
\frac{\gamma}{\gamma -1} \frac{kT}{\mu X_{\rm H} n_{\rm e}
\Lambda(T)} \label{tcool.eq}
\end{equation}
where $\gamma=5/3$ is the adiabatic index; $\mu \approx 0.61$ (for a
fully-ionized plasma) is the molecular weight; $X_{\rm H} \approx
0.71$ is the hydrogen mass fraction; and $\Lambda(T)$ is the cooling
function. We calculate the cooling function and the electron density
by following the procedure described in Sect. \ref{mass.sec}, using
the $\beta$-parameters derived by fitting the surface brightness
profile over the 65 - 1300$\;h_{70}^{-1}$\,kpc region\footnote{The
best fit obtained in Sect. 3.2.1 cannot be extrapolated to the
central region ($r \lesssim$ 230 kpc) therefore cannot be used here
for the purpose of calculating the central cooling time.} (data in
this region can be approximated by a $\beta$ model with $r_{\rm c}
\sim 113$$\;h_{70}^{-1}$\,kpc and a slope parameter $\beta \sim
0.70$). The cooling time is less than 10 Gyr inside a radius of
150$\;h_{70}^{-1}$\,kpc ($\sim 0\myarcmin6$), in agreement with the
\textit{ROSAT} result of Allen (2000).\\
We therefore accumulate the spectrum in the central $0\myarcmin6$.
We compare the MEKAL model already used in Sect. \ref{global.sec}
and \ref{radial.sec} with a model which includes a single
temperature component plus an isobaric multi-phase component (MEKAL
+ MKCFLOW in XSPEC), where the minimum temperature, $kT_{\rm low}$,
and the normalization of the multi-phase component, Norm$_{\rm low}
= \dot{M}$, are additional free parameters. The maximum temperature
$kT$ of the MKCFLOW model is linked to the ambient value of the
MEKAL model. This model differs from the standard cooling flow model
as the minimum temperature is not set to zero.
\begin{table}
\begin{center}
\caption { The best-fit parameter values and 90\% confidence limits
from the spectral analysis in the central $0\myarcmin6$ region.
Temperatures ($kT$) are in keV, metallicities ($Z$) as a fraction of
the solar value and normalizations in units of $10^{-14} n_{\rm e}
n_{\rm p} V / 4 \pi [D_{\rm A} (1+z)]^2$ as done in XSPEC (for the
MKCFLOW model the normalization is parameterized in terms of the
mass deposition rate $\dot{M}$, in $\mbox{M}_{\sun} \mbox{ yr}^{-1}
$).} \hskip 0.truein
\begin{tabular}{ccc}
\hline
Parameter   &     MEKAL    & MEKAL+MKCFLOW    \\
\hline
$kT$ & $5.2^{+0.1}_{-0.1}$ &  $8.6^{+1.1}_{-0.5}$   \\
$Z$ & $0.35^{+0.03}_{-0.02}$ &  $0.40^{+0.03}_{-0.03}$  \\
Norm & 6.67$^{+0.07}_{-0.09}\times 10^{-3}$ & 1.07$^{+0.48}_{-0.74} \times 10^{-3}$  \\
$kT_{\rm low}$ &  --- &  $1.7^{+0.2}_{-0.2}$   \\
Norm$_{\rm low}$   & --- & ${\dot M}=1580^{+150}_{-130}$ \\
$\chi^2$/dof & 2024/1783 &  1903/1781  \\
\hline
\end{tabular}
\label{cf.tab}
\end{center}
\end{table}
The results, summarized in Table\,\ref{cf.tab}, show that the
statistical improvements obtained by introducing an additional
emission component compared to the single-temperature model are
significant at more than the 99\% level according to the F-test,
with the temperature of the hot gas being remarkably higher than
that derived in the single-phase model. The fit with the modified
cooling flow model sets tight constraints on the existence of a
minimum temperature ($\sim$ 1.7 keV). We find a very high value of
the nominal mass deposition rate in this empirical model: $\sim 1600
\mbox{ M}_{\sun} \mbox{ yr}^{-1}$. \textit{ASCA-ROSAT} observations
already found a very strong cooling flow in this cluster (Allen
2000).

%%%%%%%%%%%%%%%%%%%%%%%%%%%%%%%%%%%%%%%%%%%%%%%%%%%%%%%%%%%%%%%%%%%%%%%%%%%%%%

\subsection{Mass determination}
\label{mass.sec}

\begin{figure}[ht]
\includegraphics[angle=0,width=\hsize]{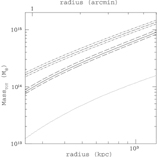}
%\special{psfile=figure/mass.ps hoffset=0 voffset=-270 vscale=40
%hscale=40 angle=0} \vspace{7.5cm}
\caption{ \textit{Solid line}:
Integrated total mass calculated from Eq. \ref{massbeta.eq} (with
error on the mass calculation coming from the temperature
measurement and $\beta$-model shown by the \textit{dashed lines}).
\textit{Dashed-dotted line }: Projected total mass with error.
\textit{Dotted line}: Gas mass. } \label{massa.fig}
\end{figure}

In the following we estimate the total mass of the cluster using the
usual assumptions of hydrostatic equilibrium and spherical symmetry.
Under these assumptions, the gravitational mass $M_{\rm tot}$ of a
galaxy cluster can be written as:
\begin{equation}
M_{\rm tot}(<r) = - \frac{kT \, r}{G \mu m_p} \left[ \frac{ d \ln
n_g}{d \ln r} + \frac{d \ln T}{d \ln r} \right] \label{mass.eq}
\end{equation}
where $G$ and $m_p$ are the gravitational constant and proton mass
and $\mu \approx 0.61$. The deprojected $d \ln n_g/d \ln r$ was
calculated from the parameters of the single $\beta$-model derived
in Sect. \ref{brightness.sec}. In particular, the advantage of using
a $\beta$-model to parameterize the observed surface brightness is
that gas density and total mass profiles can be recovered
analytically and expressed by simple formulae:
\begin{equation}
n_{\rm gas}(r) = n_{\rm 0,gas} \left[ 1+ \left(\frac{r}{r_{\rm c}}
\right)^2 \right]^{-3 \beta/2} \label{n.eq}
\end{equation}
\begin{equation}
M_{tot}(<r) = \frac{k \, r^2}{G \mu m_p} \left[ \frac{ 3 \beta r
T}{r^2 + r^2_{\rm c}} - \frac{d T}{d r} \right] \label{massbeta.eq}
\end{equation}
In estimating the temperature gradient\footnote{As a first-order
approximation, the temperature gradient is estimated by
least-squares fitting a straight line to the observed deprojected
temperature profile.} from the profile shown in Fig.
\ref{profilot.fig}, only data beyond $30''$ ($\sim
130$$\;h_{70}^{-1}$\,kpc) were considered: in the central bins the
temperature as derived in Sect. \ref{radial.sec} is more affected by
the \textit{XMM} PSF and projection effects, while for the outer
regions these effects can be neglected \citep[e.g. ][]{kaastra04a}.
The total gravitating mass distribution derived from Eq.
\ref{massbeta.eq} is shown in Fig. \ref{massa.fig} as a solid line,
with errors coming from the temperature measurement and
$\beta$-model represented as dashed lines. Within 1 Mpc we find a
total mass of $\sim (5.4 \pm 0.5) \times 10^{14} M_{\sun}$ and
within the outer radius of the cluster as visible in the X-ray
surface brightness profile (1.5 Mpc) we find $\sim (8.5 \pm 0.7)
\times 10^{14} M_{\sun}$. We note that the total integrated mass
within a particular volume is dependent upon the local physical
properties (temperature and density gradients) and is unaffected by
the regions interior, or exterior, to that radius. The mass profile
derived with this method is thus reliable in the region where the
$\beta$-model is a good representation of the observed surface
brightness profile (230 kpc $\lesssim$ \, r \, $\lesssim$ 1300 kpc,
see Sect. 2.3.1), whereas it cannot be extrapolated to the central
region.\\
We also calculate the projected mass along the line-of-sight within
a cylinder of projected radius $r$. The integration was performed
out to a radius of $\sim 5$ Mpc from the cluster centre. The
projected total mass is shown in Fig. \ref{massa.fig} as a
dashed-dotted line.
~\\
In Fig. \ref{massa.fig} we also show (dotted line) the gas mass
profile derived by integrating the gas density given by Eq.
\ref{n.eq} in spherical shells and using the $\beta$-model
parameters determined in Sect. \ref{brightness.sec}. The
normalization of Eq. \ref{n.eq} is obtained from the combination of
the best-fit results from the spectral and imaging analyses, which
allows us to determine the conversion count rate - flux used to
derive the bremsstrahlung emissivity that is then integrated along
the line-of-sight and compared with the central surface brightness
value. We note that, since we adopt the parameters of the
$\beta$-model fit in the outer regions, the derived central electron
density ($n_{\rm 0,e} \sim 1.7 \times 10^{-2} \mbox{ cm}^{-3}$) is
that predicted by the extrapolation of the $\beta$-model fit to the
centre (see Fig. \ref{1betafit}). This procedure is nonetheless
reliable in estimating the gas mass for $r >
230$$\;h_{70}^{-1}$\,kpc, shown in Fig. \ref{massa.fig}.\\
In order to allow a direct comparison with our weak lensing studies
in Sect.\,\ref{subsect:wl} and to derive an estimate of $r_{200}$ we
perform a fit to the NFW profile \citep{nfw1,nfw2} given by
\begin{equation}
M_{\rm DM} (<r) = 4 \pi r^3_{\rm s} \rho_{\rm c} \frac{200}{3}
\frac{c^3 \left( \ln (1+r/r_{\rm s}) - \frac{r/r_{\rm s}}{1 +
r/r_{\rm s}} \right)}{ \ln (1+c) - c/(1+c)} \label{nfw.eq}
\end{equation}
where $\rho_{\rm c}$ is the critical density. The scale radius
$r_{\rm s}$ and the concentration parameter $c$ are the free
parameters. The best fit parameters that minimize the $\chi^2$ of
the comparison between the mass predicted by the integrated NFW dark
matter profile and the mass profile reconstructed from Eq.
\ref{massbeta.eq} are: $r_s = 516.6 \pm 13$ kpc and $c = 3.58 \pm
0.07$, where the relation $r_{200} = c \times r_s$ holds. The quoted
error are at the 68\% confidence levels (1$\sigma$) on a single
parameter of interest. Note that we neglect the gas mass
contribution to the total mass and we assume $ M_{\rm tot} (<r) =
M_{\rm DM} (<r) $. We have also performed the same fitting procedure
by including the gas mass, i.e. by assuming $ M_{\rm tot} (<r) =
M_{\rm DM} (<r) + M_{\rm gas} (<r) $, and find very similar results.
In Fig.\,\ref{fig:nfwcontours}(a) we show confidence contours of the
NFW fit to our mass profile and note that both parameters are well
constrained. For a comparison to the equivalent model based on
lensing data we refer to Sect.\,\ref{sec:summary}.
%

%
%_______________________________________________________________________________________________________________________________________
%_______________________________________________________________________________________________________________________________________
\section{Optical observations}
\label{sec:opt_observations}
\subsection{WFI-observations and data reduction}
\begin{figure}
   \centering
   \centerline{\includegraphics[angle=0,width=\hsize]{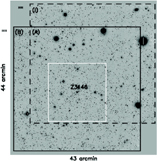}}
   \caption{Layout of the Z3146 $R$-band WFI observations
     of the Bonn (B; solid lines) and Innsbruck (I; dashed lines)
     groups. The orientation with respect to the sky is North-Up and
     East-Left.  The overlap in the $R$ exposures allows useful tests
     in the shape measurement procedures during our weak lensing
     analysis (see Sect. \ref{subsect:wl}). If not stated otherwise we
     use the deep $R$ band stack from both observing campaigns and
     work in the common area (A) where we have data in $B$, $V$ and
     $R$ available. It has an extent of $25\myarcmin8\times
     23\myarcmin 8$ and the cluster is located $8\myarcmin0$ north
     and $10\myarcmin 0$ west with respect to its lower left corner. The white
     square marks the $\sim16\arcmin40\arcsec\times16\arcmin40\arcsec$ region used in Sect.\,\ref{sec:stronglensing} and
     \ref{sec:furtherinvestigations} for extracting the Red Sequence galaxies.}
         \label{fig:layout}
\end{figure}

\begin{figure}
   \centering
   \centerline{\includegraphics[angle=0,width=8cm]{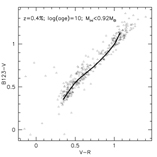}}
      \caption{As a control check for our photometric calibration
      we plot colours of bright, unsaturated stars in our WFI fields
      compared to isochrones for stars of 10 Gyrs with a metallicity
      of 0.4\% \citep[see][]{girardi02a}. Magnitudes for the stars
      were estimated
      with the SExtracor MAG\_AUTO magnitude. Our data are in excellent agreement
      with the isochrone predictions.}
         \label{fig:photometry}
\end{figure}

Z3146 was observed with WFI@ESO/MPG2.2m in the two observing
programs 68.A-0255 (P.I. S. Schindler) and 073.A-0050 (P.I. P.
Schneider). The first one obtained 8000s in broad band $V$
(BB\#V/89\_ESO843) and 16100s in broad band $R$ (ESO filter
BB\#Rc/162\_ESO844). The second programme observed for another 8900s
in $R$ and 1500s in $B$ (BB\#B/123\_ESO878).  All data were
processed with our image reduction pipeline developed within the
GaBoDS project. It performs all necessary steps from raw frames to
astrometrically and photometrically calibrated and co-added images.
The individual methods and its performances on WFI data are
described in detail in \cite{schirmer03a} and \cite{pipeline1}. All
data were obtained during clear nights, under good seeing conditions
($\le 1\myarcsec2$) and with a large dither box of about
$3\myarcmin0$ to ensure good flat-fielding and an accurate
astrometric calibration. The images were tied to the astrometric
frame of the USNO-A2 catalogue \citep{mon98}, photometrically
calibrated with Stetson standards \citep{stetson00a} and finally
co-added with the {\tt Swarp} tool \citep{swarp}. We produced
several co-added images from our $R$ band exposures mainly to
crosscheck our object shape measurements in the weak lensing
analysis (see Sect. \ref{subsect:wl}). The characteristics of all
co-added images used in this work are summarised in
Table\,\ref{tab:data} and Figs. \ref{fig:layout} and
\ref{fig:photometry}. Each co-added science image has a pixel scale
of $0\myarcsec238$ and is accompanied by a weight map characterising
its noise properties.

\begin{table*}
\begin{center}
\begin{tabular}{ccccc}
\hline \hline
Filter & Image Region & exp. time (s) & limiting magnitude ($3\sigma$ in a $2\myarcsec 0$ aperture) & Seeing (arcsec)\\
\hline
$B$ & (B) & 1500 & 24.70 & 1.21 \\
$V$ & (I) & 8040 & 25.02 & 1.16 \\
$R$ & (I) & 16077 & 25.27 & 1.02 \\
$R$ & (B) & 8850 & 25.24 & 1.11 \\
$R$ & (A) & 24927 & 25.71 & 1.04 \\
\hline
\end{tabular}
\end{center}
\caption{Characteristics of the co-added WFI images. The limiting
magnitudes quoted in column 4 are defined in the Vega system via
$m_{\rm lim}=ZP-2.5\log(\sqrt{N_{\rm pix}}\cdot 3\cdot\sigma)$,
where $ZP$ is the magnitude zeropoint, $N_{\rm pix}$ is the number
of image pixels in a circle with radius $2\myarcsec 0$ and $\sigma$
the sky background noise. The seeing in column 5 was measured with
the {\tt SExtractor} FWHM\_IMAGE parameter. The groups in Bonn (B)
and Innsbruck (I) observed Z3146 with an offset of about $5\myarcmin
0$ In Ra and $8\myarcmin 0$ in Dec. The exact layout is given in
Fig. \ref{fig:layout}. For the $R$ band we created, besides the
individual stacks, a deep mosaic in the common area (A). We note
that the Innsbruck $R$ band has nearly double the exposure time of
the Bonn image but about the same limiting magnitude as it was
observed during less favourable moon phases. The quality of our
photometric calibration is crosschecked in Fig.
\ref{fig:photometry}.\label{tab:data}}
\end{table*}

%--------------------------------------------------------------------------
%--------------------------------------------------------------------------
%--------------------------------------------------------------------------
%--------------------------------------------------------------------------

\subsection{HST-observations}
In addition to our WFI observations we used archival calibrated HST
data from the WFPC2 Associations obtained during a snapshot
programme (PI: Edge, PID-number 8301). Z3146 was observed with the
WFPC2 (filter F606W) in April 2000 for a total exposure time of
1000\,s. The pixel scale is $\sim0\myarcsec 1$ per pixel, the seeing
is measured with the \tt FWHM\_IMAGE \rm keyword of \tt SExtractor
\rm to be $\sim0.1\arcsec$. The main purpose using these archival
HST data was the identification and
comparison of arc candidates with the ground based observations.\\

%_______________________________________________________________________________________________________________________________________
\section{Lensing analysis}
\label{sec:opt_analysis}
\subsection{Weak lensing analysis}
\label{subsect:wl} As a second method for determining the mass and
its distribution of Z3146 we performed a weak lensing mass
reconstruction. For a broad introduction on weak lensing and its
application in cluster mass determinations see for instance
\cite{weaklensing}. In the following we describe the creation of our
background galaxy catalogue and the weak lensing analysis.
Throughout the analysis we use standard weak lensing notation.
\subsubsection{Lensing catalogue generation}
We use the deepest image, the $R$-band for the weak lensing
measurement.  To create an object catalogue with shear estimates for
all objects we first extract sources with the {\tt SExtractor} (we
use a detection threshold of 1.9 and a minimum area of 3 pixels for
our detections) and the {\tt IMCAT\footnote{see
http://www.ifa.hawaii.edu/$\sim$kaiser/imcat}} softwares. While {\tt
SExtractor} produces a very clean object catalogue if the source
extraction from the science images is supported by a weight map
\citep[see e.g. Fig. 27 of][]{pipeline1}, {\tt IMCAT} calculates all
quantities necessary to estimate object shapes. We merge the two
catalogues and calculate shear estimates with the KSB algorithm
\citep[see][]{kaiser95a}. For the exact application of the KSB
formalism we closely follow the procedures given in \cite{erben01a}
with important modifications in the selection of stars that are used
for the necessary PSF corrections \citep[see][]{vanwaerbeke05a}.
After the PSF corrections we reject all objects with an {\tt IMCAT}
significance $\nu<8$, a half light radius smaller or equal to that
of bright stars, and a final modulus of the shear estimate $|e| >
0.8$. In the following analysis we do not apply a weighting to
individual galaxies.

\begin{figure}
   \centering
   \centerline{\includegraphics[angle=0,width=\hsize]{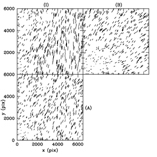}}
      \caption{Stellar anisotropies for the fields (I), (B) and (A)
      in their common area.The length of the bars gives a measure for the
      stellar ellipticity and the direction the orientation of the semi-major
      axis. The longest bars correspond to an ellipticity of about $5\%$.
      All patterns show a smooth variation over the whole field-of-view of
      the (A) area.}
         \label{fig:PSFellip}
\end{figure}

\begin{figure}
   \centering
   \centerline{\includegraphics[angle=0,width=\hsize]{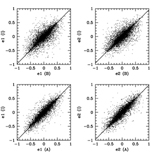}}
   \caption{Galaxy ellipticities after all PSF corrections
     from completely independent analyses of the (I), (B) and (A)
     images in the overlap area. In the upper panels we compare the
     (B) and (I) sets which are independent (in terms of raw images
     entering the co-added stacks) from each other. The slopes of the
     fitted lines are 0.983 for e1 and 0.95 for e2. For the lower
     panels (comparison between (A) and (I)) the slopes are 0.992 for
     e1 and 0.974 for e2. The corresponding numbers for (A)-(B) are
     1.080 for e1 and 1.091 for e2 (not shown in the figures).
     We conclude that the measurements are in
     very good agreement with each other. For the fits, the Ordinary
     Least Squares bisector method \citep[see][]{isobe90a},
     which considers errors in e1 and e2, is chosen.
     After applying the line fits to the ellipticity components
     the mean shear differences are compatible with zero in
     all four cases.
     We note that the
     standard deviation of the residuals $\sigma_{\rm \Delta e}$ is
     0.16 in both components for the (B)-(I) comparison but 0.10 in
     the (A)-(I) case. The cause for the lower value in the (A)-(I)
     case is that the data are not independent but share a large fraction
     of the input images.
     The residuals as function of position
     are discussed in Fig. \ref{fig:ellipcomppos}.}
         \label{fig:ellipcomp}
\end{figure}

\begin{figure}
   \centering
   \centerline{\includegraphics[angle=0,width=\hsize]{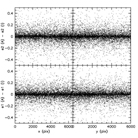}}
      \caption{Differences of corrected galaxy ellipticity measurements
      from the (I) and (A) images as function of image position.
      We see no systematic trends of residuals
      with object position. Also a closer, visual inspection of the two-dimensional
      ellipticity residual distribution reveals no systematics on small
      scales. The standard deviation of the ellipticity components around
      zero is $\sigma_{\rm \Delta e}\approx 0.1$ in all four cases.
      Hence we conclude, together with the results
      of Fig. \ref{fig:ellipcomp}, that it is safe to use the (A) image
      for our weak lensing analysis.}
         \label{fig:ellipcomppos}
\end{figure}

Because of the large offsets between the observations from the Bonn
and Innsbruck groups it is not obvious whether we can savely use the
deep stack (A) for the lensing analysis or whether we have to work
on the individual co-additions (I) and (B).  This mainly comes from
the assumption of a smooth variation of the PSF over the whole
field-of-view when we correct galaxy shapes for PSF effects. With
the large offsets we could suffer from discrete jumps in the PSF
anisotropy within the (A) mosaic. However, in Fig.
\ref{fig:PSFellip} we see that the PSFs of all three stacks are well
behaved in the (A) area.  We performed comparisons of the final
shear estimates in the (I), (B) and (A) mosaics and found that they
are in excellent agreement (see Figs. \ref{fig:ellipcomp} and
\ref{fig:ellipcomppos}).  Therefore we decided to use the (A)
mosaic, which is the deepest image, for our primary analysis.

\begin{figure}
   \centering
   \centerline{\includegraphics[angle=0,width=\hsize]{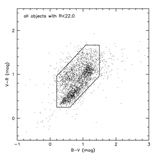}}
      \caption{To obtain criteria to reject probable low $z$ foreground
               objects from our lensing analysis we plot all bright objects with $R<22.0$
               in a colour-colour diagram. Most of them lie in a well defined area which we
               use to clean our catalogue. Elliptical cluster galaxies for a $z=0.3$
               cluster are expected around $B-V\approx 1.2;\; V-R\approx 1.2$.}
         \label{fig:colcol}
\end{figure}

The next step is to clean our lensing catalogue from likely cluster
members, foreground galaxies and faint stars. To this end we plot
stars and bright galaxies ($R<22.0$), which have a high probability
to be at a lower redshift than the cluster, in a colour-colour
diagram (see Fig. \ref{fig:colcol}). We note that all these objects
mostly occupy a limited and well defined area in colour-colour
space. We finally use the following criteria to clean our background
source catalog: We reject all objects with $R<22.0$ and keep objects
between $22.0<R<23.0$ if they do not lie in the following area of
the $(B-V)$ vs. $(V-R)$ diagram: $-0.23<(V-R)-0.8(B-V)<+0.8; \;
0.25<(V-R)<1.67; \; 0.2<(B-V)<1.5$. We keep all
objects with $R>23.0$ as probable background sources. We note that
\cite{clowe02a} and \cite{dietrich05a} used similar criteria to
identify foreground objects.

For our final lensing catalogue we additionally exclude all objects
falling in masked image regions (around bright stars, satellite
tracks etc.). This leaves us with about 12 galaxies per sq. arcmin
as direct tracers of the cluster shear. Around the Brightest Cluster
Galaxy (BCG) of Z3146, this average density is reached at a radius
of about $45\myarcsec 0$ with only very few sources in the cluster
centre.

%--------------------------------------------------------------------------
%--------------------------------------------------------------------------
%--------------------------------------------------------------------------
%--------------------------------------------------------------------------
%--------------------------------------------------------------------------
%--------------------------------------------------------------------------
%--------------------------------------------------------------------------
%--------------------------------------------------------------------------
%
\subsubsection{Weak lensing mass determination of Z3146}
\label{sec:lensmass} 
\begin{figure*}
   \centering
   \centerline{\includegraphics[angle=-90,width=0.45\hsize]{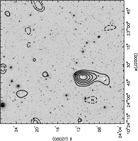}
   \includegraphics[angle=-90,width=0.45\hsize]{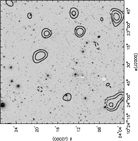}}
   \caption{KS93 weak lensing cluster mass reconstruction of Z3146
     (left panel). The shear field was smoothed with a Gaussian of width
     $1\myarcmin 4$. The $\kappa$ map was normalised so that the mean of
     $\kappa$ in the entire field is zero. Because we use this map only to define
     the centre of mass for model fits to the shear data the actual choice for this
     normalisation is not important. Solid contours increase in steps of
     $\Delta\kappa=0.01$ $(\sim 4.29\times 10^{13}M_\odot /$Mpc$^{2}$),
     assuming a source galaxy redshift of 0.79. The cluster clearly shows up
     as the highest peak in this reconstruction. Dashed contours represent the
     same levels of negative $\kappa$.  The white cross marks the cluster
     centre of our X-ray analysis (see Sect.  \ref{morphology.sec}) which is
     in excellent agreement with the lensing centre. The peak at the cluster
     position is detected with a significance of more than 4.1$\sigma$, the
     extension to the South (marked with a white circle) with
     $2.7\sigma$. Hence, this extension of the $\kappa$ map is not considered
     a significant lensing feature. The right panel shows a B-mode map of the
     $\kappa$ field. It contains two deeper holes in the southern part that
     are significant with $2.9\sigma$ (eastern hole) and $3.5\sigma$ (western
     hole). Around the cluster position this map shows a null signal.  See the
     text for more details.}
         \label{fig:massreco}
\end{figure*}

\begin{figure}
   \centering
   \centerline{\includegraphics[angle=0,width=0.9\hsize]{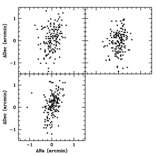}}
   \caption{The lower left plot shows the lensing centres of our
     cluster reconstructions after adding noise realisations to our
     signal (see text for details). The quoted distances are with
     respect to the original cluster centre. Formal $1\sigma$
     positional uncertainties of the lensing peak derived from this
     distribution are $\Delta R.A.=19\myarcsec 8$ and $\Delta
     Dec=45\myarcsec 6$.  We investigated whether the observed
     asymmetry of the error distribution comes from shot noise in the
     galaxy ellipticities (the error might be dominated by a few very
     elliptical galaxies close to the cluster centre) or from the
     galaxy positions (as areas around bright stars or other image
     defects have been masked out the object distribution in our field
     is not homogeneous). To this end we assigned each galaxy a random
     ellipticity of a Gaussian distribution with $\sigma_e=0.38$ (see
     Sect. \ref{sec:lensmass}) and repeated our error analysis with
     200 new noise realisations from this catalogue (upper left plot).
     For the upper right plot we additionally randomised the positions
     of our objects.  Both simulations show a similar asymmetry as the
     original analysis and we conclude that it does not originate from
     shot noise of the galaxies. We note that the positional error estimates of this
     analysis are an upper limit as the original signal is already
     noisy (we implicitly assumed it to be noise free in our calculations).}
         \label{fig:posdiff}
\end{figure}

The main interest of the weak lensing analysis
in this work is an estimate of the total cluster mass of Z3146 and
its inter-comparison with the X-ray analysis. We first perform a
standard KS93 cluster mass reconstruction \citep[see][]{kaiser93a}
to investiagte the dark matter distribution and to obtain an
estimate for the cluster centre. In addition, we calculate a {\sl
B-mode} map, i. e.  we performed another mass reconstruction after
all object ellipticities have been rotated by 45 degrees. This map
should contain noise only if the lensing data are free from
systematics. The results are discussed in Fig. \ref{fig:massreco}.
We see that our lensing centre is in excellent agreement with that
determined from our X-ray analysis (see also Sect.
\ref{morphology.sec}) and we use the X-ray position for the
following analysis. We estimate significances for peaks in our
reconstructions and errors on the lensing centre with the following
procedure: We randomise the orientation of our galaxies, redo a KS93
mass reconstruction with the new catalogue and repeat this procedure
many times.  For the peak significance we count how often the
$\kappa$ value in our noise maps exceeds that of the lensing signal.
With 29700 realisations the probabilities that the cluster peak, the
cluster mass extension and the eastern and western holes in the
B-mode map are pure noise features are 0/29700, 170/29700, 88/29700
and 9/29700.  Assuming Gaussian statistics this translates to
significances of $>4.1\sigma$, $2.7\sigma$, $2.9\sigma$ and
$3.5\sigma$ respectively. We conclude that the highly significant
cluster peak has no significant extension to the South. Next we
estimate errors on the lensing peak position. The best way to
measure the centroid dispersion would be to use a parametric model
for the mass concentration and to generate noisy mass maps with
randomised ellipticity orientations and galaxy positions. If the
model were true we would obtain accurate error estimates from the
noisy mass realisations. With the data at hand we can follow this
idea by considering the original reconstruction as the input mass
model. We probably overestimate the true error in this way because
our input model already contains measurement noise. We plot the
result of this exercise for 200 maps in the lower left panel of Fig.
\ref{fig:posdiff}. We find a significant asymmetry in the
distribution of positional differences. The positional accuracy is
about 2.3 times better along R.A. than in Dec. As the mass
distribution is elongated towards the South we would expect a skewed
distribution of the positional errors towards negative Dec values
but the observed symmetric elongation in the North-South direction
is surprising.  We checked that not a few, very elongated galaxies
or shot noise from the galaxy positions (introduced for instance by
object masks) are responsible for this result (see Fig.
\ref{fig:posdiff}). Given this result we quote the positional
accuracy of our lensing centre as $\Delta {\mbox R\rm a}=19\myarcsec
8; \; \Delta {\mbox D\rm ec}=45\myarcsec 6$.

\begin{figure}
   \centering
   \centerline{\includegraphics[angle=0,width=0.9\hsize]{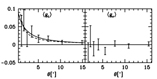}}
      \caption{{\sl Left:} Tangential shear signal as function
      of radius from the centre of Z3146. The signal is robust and we can trace
      the cluster shear up to the border of our data field. The solid line
      shows the reduced shear of our best fit NFW model $(r_{200}=1661\;h_{70}^{-1}\mbox{kpc}, \,
      c=3.6)$, the dashed line that of an SIS fit ($\sigma_v=869$km/s).
      See the text for further details. {\sl Right:} We show the cross component
      of the shear $g_r$ around the cluster centre, i. e. the signal after
      rotating all galaxies by 45 degrees. It should vanish if lensing caused
      the original signal and our measurement is compatible with zero over the
      entire distance range (see also Fig. \ref{fig:massreco}). As
      a further consistency check we compare in Fig.
      \ref{fig:tangcomp} tangential shear measurements from the (I), (B) and
      (A) data sets.}
         \label{fig:tangg}
\end{figure}

While our mass maps give us insight into the dark matter
distribution in Z3146 it is difficult to obtain reliable estimates
for the total lensing mass and the involved errors. The main
problems are that mass reconstructions involve a convolution from
the measurable shear field and, in addition, they become very noisy
at a distance several arcminutes from the cluster centre (see Fig.
\ref{fig:massreco}). Moreover, they intrinsically suffer from the
mass-sheet degeneracy. Hence, we will estimate a lensing mass by
directly fitting parametrised lensing models to the shear data.  The
error analysis is simplified significantly in this case. Moreover,
model fits break the mass-sheet degeneracy by the explicit
assumption that $\kappa$ at large distances from the lensing centre
is zero. The main drawback is that shear data alone do not allow a
clear discrimination between different, plausible mass models (see
Fig. \ref{fig:tangg}).

For our model fits to the shear data we primarily consider the
universal density profile (NFW) proposed by \cite{nfw1}. The details
for the calculation of the lensing quantities $\kappa$ and $\gamma$
for this profile are given in several publications and the details
are not repeated here \citep[see e.g.][]{bartelmann96a,kruse99a}. To
determine our model parameters we use the log-likelihood method
proposed in \cite{schneider00a}. We maximise the likelihood
function:
\begin{equation}
   \label{eq:shearlikelihood}
   {\cal L}=\prod_{i=1}^{N}{1\over \pi\sigma^2[\boldg(\theta_i, \bolda)]}
   \exp\left(-{|\bolde_i+\boldg(\theta_i, \bolda)|^2 \over \sigma^2[\boldg(\theta_i, \bolda)]}\right) \;,
\end{equation}
where $N$ is the number of galaxies, $\bolde$ the observed,
two-dimensional ellipticity of each galaxy, $\boldg$ the reduced
shear of the model, $\bolda$ the set of parameters to be fitted,
$\theta$ the galaxy position in the lens plane and $\sigma$ the
dispersion of the observed ellipticity. It is given by
$\sigma=(1-|g|^2)\sigma_{e_{\rm s}}$, where $\sigma_{e_{\rm s}}$
stands for the (two-dimensional) dispersion of the unlensed source
ellipticities. All the model parameters $\bolda$ are contained in
$\boldg$. The fundamental assumption of this method is that the
source ellipticity distribution can well be described by a Gaussian
distribution of width $\sigma_{e_{\rm s}}$. It is optimal in the
sense that it uses the full ellipticity information and not only
individual components (such as fits to the tangential part of the
shear). For a more detailed discussion on this likelihood method and
its properties see \cite{schneider00a}.

Before we apply this method we still have to specify the redshifts
of the source galaxies and the galaxy sample we include in our fits.
In Fig. \ref{fig:tangg} we show the tangential component of the
shear around the cluster centre. We can trace the cluster shear
signal over the whole field-of-view of $15\myarcmin 0$\footnote{We
note that for radii larger than $8\myarcmin 0$ we cover the area
around the cluster completely only North-West of its centre (see
Fig. \ref{fig:layout})}.  Hence, we include all preselected
background galaxies in our estimations. For the dispersion of the
unlensed ellipticities $\sigma_{e_{\rm s}}$ we use the measured
value of our galaxies $\sigma_{e}$ averaged over the whole
field-of-view. Here we assume that weak lensing does not change this
value significantly and we estimate $\sigma_{e_{\rm s}}=0.38$. For
the redshift distribution of our background galaxies we use
estimates from \cite{hss06}. The authors obtained photometric
redshifts for 62000 galaxies with $21.5<R<24.5$. Their WFI data
consist of 1.75 sq. degree of deep $UBVRI$ photometry in three
different patches \citep[see][for details on the data]{hss06,hed06}.
The photometric redshift distribution is parametrised by the
following function introduced by \citet{blp06}:
\be p_{\rm fit}(z)=A\,\left[p_1(z)\,H(z_{\rm
t}-z)+p_2(z)\,H(z-z_{\rm t} )\right], \label{redshiftdistribution}
\ee
where $H$ denotes the Heaviside step function,
\be p_1(z) = \left(\frac{z}{z_0}\right)^\alpha
\exp\left[-\left(\frac{z}{z_0}\right)^\beta\right] \ee
and
\be p_2(z) = \exp{\left[\left(\frac{z_t}{z_1}\right)^\gamma -
\left(\frac{z}{z_1}\right)^\gamma \right]}\,p_1(z_{\rm t}); \ee
the normalisation $A$ is obtained by
\be \int_0^{z_{\rm t}} \dd z\, p_1(z) + \int_{z_{\rm t}}^{\infty}
\dd z\, p_2(z) = 1. \ee
For $z_{\rm t}=1$ the reported fit parameters are:  $z_0=0.27$,
$\alpha=2.84$, $\beta=1.40$, $\gamma=2.34$ and $z_1=2.16$. With this
choice the mean redshift for galaxies behind our cluster at $z_{\rm
cl}=0.29$ is $\langle z_{\rm bg}\rangle =0.79$. We use this
distribution to estimate the average geometrical lensing factor: \be
D_{\rm d}(z_{\rm cl})\left\langle \frac{D_{\rm ds}(z_{\rm cl},
z)}{D_{\rm s}(z)} \right\rangle =
 D_{\rm d}(z_{\rm cl})\int_{z_{\rm cl}=0.29}^{\infty} \dd z\, p_{\rm fit}(z) \frac{D_{\rm ds}(z_{\rm cl}, z)}{D_{\rm s}(z)},
\ee where $D_{\rm d}$, $D_{\rm ds}$ and $D_{\rm s}$ represent
angular diameter distances observer-cluster, cluster-background
source and observer-background source respectively.
%
%{\bf To get an idea for the redshift distribution
%of our background sources we use photometric redshifts estimated from
%the Chandra Deep Field South \citep[see][]{wolf04a}.
%For the galaxy sample in the magnitude range $22<R<24$
%(the authors quote $R=24$ as their faint limit for reliable $z$
%estimates) and $z>0.29$ their
%distribution has a mean of $z_s\approx 0.75$ and in the following
%we put all source galaxies to that redshift.} If we instead used a
%$z_s$ of unity, our final mass estimates would be lowered by
%$10\%-15\%$ (see also Fig. \ref{fig:lensmass}).

For our fits to the NFW profile, we consider the concentration $c$
and the radius $r_{200}$ \citep[see][]{nfw1} as free model
parameters. With our setup, the application of our prescription to
the shear data leads to best fit values of
$r_{200}=1661^{+280}_{-328}\;h_{70}^{-1}$kpc and
$c=3.6^{+2.8}_{-2.4}$. The model has a significance of 4.35 over one
with zero mass and the errors on $r_{200}$ and $c$ are at the $90\%$
confidence level. They were estimated with our likelihood analysis
by keeping $c$ or $r_{200}$ at its best fit value and leaving
$r_{200}$ or $c$ as the only free parameter. In Fig.
\ref{fig:nfwcontours}(b) we show confidence contours of our analysis
and note that both parameters are reasonably well constrained except
for low values of $c$ see Sect.\,\ref{sec:summary} for a comparison
to an NFW model based on X-ray data). In addition to the NFW profile
we also modeled our shear data by a Singular Isothermal Sphere (SIS)
characterised by its velocity dispersion $\sigma_v$. Our best fit
model has $\sigma_v=869^{+124}_{-153}$km/s. The errors represent the
90\% confidence level and the model has a significance of
$4.88\sigma$ compared to the zero mass model. In contrast to the NFW
fit, the significance and the estimated velocity dispersion of the
SIS model show some dependence on the galaxies included close to the
cluster centre. We notice an increase of significance by excluding
the galaxies in a circle $30\myarcsec 0$ around the cluster (five
objects) and a smooth decrease of $S/N$ if we reject more galaxies
beyond that point. Hence, we used all galaxies with a distance
greater than $30\myarcsec 0$ from the cluster centre for our SIS
fit.

We finally discuss a possible bias of our result due to a systematic
underestimate of the shear. As we showed in \cite{erben01a} and
within the Shear Testing Program \citep[see][]{heymans05a} our
pipeline may underestimate weak shear by 10\%-15\%.  We recalculated
the best fit NFW values after boosting all ellipticities by a factor
of 1.15. We then obtain $r_{200}=1701^{+261}_{-303}\;h_{70}^{-1}$kpc
and $c=3.66^{+2.52}_{-2.18}$ which is well within the error bars of
the original signal. Hence, a possible systematic underestimate of
the shear by about 15\% would not change our results significantly.
At the end of this section we show in Fig. \ref{fig:lensmass} the
total mass properties given by our model fits. We also present a
mass-to-light ratio analysis in Sect.\,\ref{subsubsec:masstolight}
and will compare our results with masses from X-ray analyses in
Sect.\ref{sec:summary}.
%--------------------------------------------------------------------------
%--------------------------------------------------------------------------
%--------------------------------------------------------------------------
%--------------------------------------------------------------------------
%--------------------------------------------------------------------------
%--------------------------------------------------------------------------
%--------------------------------------------------------------------------
\begin{figure}
   \centering
   \centerline{\includegraphics[angle=0,width=0.9\hsize]{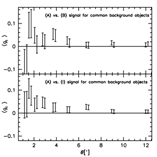}}
      \caption{The figure shows comparisons of the tangential shear signal
      around Z3146 from common background objects for the (A) and (B)
      (upper panel) and the (A) and (I) sets (lower panel). The left
      measurements are from (A) and the shifted ones from (B) and (I)
      respectively. The catalogues
      of common sources were created by merging our lensing objects
      from (A) with the ellipticity catalogues from (I) and (B); see
      Fig. \ref{fig:ellipcomp}. As in the ellipticity comparisons,
      the tangential shear signals around the cluster agree reasonably well.
      A very different behaviour is observed for the first bin to which only
      about a dozen galaxies contribute.
      Note that the initial catalogues were created independently and hence
      the object samples in the two comparisons are not exactly the same.}
         \label{fig:tangcomp}
\end{figure}
%--------------------------------------------------------------------------

%--------------------------------------------------------------------------
\begin{figure}
   \centering
   \centerline{\includegraphics[angle=0,width=0.9\hsize]{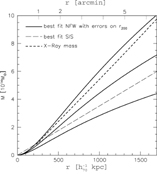}}
      \caption{Total weak lensing mass calculated from our best fit mass models
      in spheres of radius $r$ around the cluster centre. Solid lines encompass
      our best fit NFW model $r_{200}=1661^{+280}_{-328}\;h_{70}^{-1}$kpc; $c=3.6$
      with the errors from the measurement in $r_{200}$. With this
      model our total mass at $r_{200}$ is $M_{200}=7.05^{+2.8}_{-2.72}\times10^{14}M_{\odot}$.
      The long-dashed curve represents our best SIS fit to the shear data. The
      mass at $r_{200}$ is 17\% lower in this case. The total X-Ray mass of
      Z3146 (see Sect.\,\ref{mass.sec}) is given by the short-dashed line. It is
      $M_{200}=9.57^10{14}M_{\odot}$.
         \label{fig:lensmass}
   }
\end{figure}

%_______________________________________________________________________________________________________________________________________
\subsection{Strong lensing analysis}\label{sec:stronglensing}
\subsubsection{Definition/identification of gravitational arc candidates}\label{subsec:identifying}
\begin{figure*}
   \centering
   \includegraphics[angle=0,width=18cm]{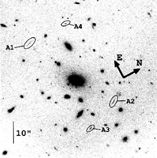}
      \caption{HST image of Z3146 obtained by the WFPC2. The exposure time was 1000 s using the filter F606W.
      The objects \tt A3 \rm and \tt A4 \rm were found on the HST frame
      by visual inspection (see Sect.\,\ref{subsec:identifying} for more details).}
         \label{fig:HST_arcs}
\end{figure*}

\begin{figure*}
   \centering
   \includegraphics[angle=0,width=18cm]{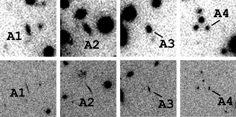}
      \caption{Detailed comparison of the arc candidates in Z3146 (top: WFI,
      bottom: WFPC2). In all images North is up, East to the left, the FoV is $15\arcsec\times15\arcsec$. See text for more details.}
         \label{fig:arcdetail}
\end{figure*}

In addition to the weak lensing analysis we have searched the
central cluster region for strongly lensed objects. In ground based
observations usually only arcs tangentially aligned with respect to
the mass centre are visible, as radial arcs are very thin and faint
structures in the vicinity of bright central galaxies of clusters.
In addition, arcs and their counter images have the same spectra and
redshifts $\gtrsim2\times z_{\rm lens}$. However we do not have
spectra, hence apart from the position and the morphology, the
redshift is the main identification criterion. Therefore we
investigated whether it is possible using our observations to
roughly estimate the photometric redshift or at least to find out
whether an object belongs to a fore- or background population. For
that purpose we have performed simulations using the software
package \tt hyperz\rm\footnote{\tt\scriptsize
http://webast.ast.obs-mip.fr/hyperz} \citep{hyperz}. We created a
set of 3000 artificial galaxies with the following parameters:
$0\leq z_{sim}\leq2$, $R$ magnitude $22\leq m_R\leq 25$ (which
corresponds to the range of the arc candidates), using the simulated
filter WFI $R$ band (ESO844) as the reference filter. The type of
the galaxies was also randomly chosen to be either E, S0, Sa, Sb, Sc
or Sd. The simulations have shown that it is not possible to obtain
any reliable redshift estimate from $BVR$ images only. 38\% of all
simulated galaxies with $z\leq z_{Z3146}$ were found to be
background objects. On the other hand, 27\% of the background
galaxies (defined as $z\geq z_{Z3146}$) were measured to be located
in front of Z3146. Hence it is even not possible to decide whether
an object of unknown redshift is a foreground or a background object
and we have to restrict our search
for strongly lensed objects to morphological criteria only.\\[2mm]
Unfortunately there is no common definition of an arc candidate. The
definition we adopt of a gravitational arc candidate is that of an
elongated object, aligned tangentially with respect to the cluster
center, a minimum length of 1\myarcsec 0 and a length-to-width ratio
$l/w\geq1.5$. However it is not yet clear whether Z3146 can produce
strong lensing or not: the low concentration parameters $c$ obtained
during the modelling of both, the X-ray ($c = 3.58 \pm 0.07$,
$r_{200}=1849\pm 47\;h_{70}^{-1}$kpc, see Sect.\,\ref{mass.sec}) and
the weak lensing data ($c=3.6^{+2.8}_{-2.4}$,
$r_{200}=1661^{+280}_{-328}\;h_{70}^{-1}$kpc, see
Sect.\,\ref{sec:lensmass}) to an NFW profile, leads to an Einstein
ring of only $\sim1\arcsec$. However, due to the large errors in
both the $c$ and $r_{200}$ determination and the unknown source
redshift, we cannot exclude the strong lensing ability: a source
redshift of $z=2$ and adopting the upper limits of $c$ and $r_{200}$
(leading to $c=6.4$ and $r_{200}=1941\;h_{70}^{-1}$kpc, based on
weak lensing values) shifts the Einstein ring to $\sim13\arcsec$.
Additionally, our adopted SIS model derived in
Sect.\,\ref{sec:lensmass} leads to an Einstein radius of the same
size, assuming the source located at the derived mean redshift
$\langle z_{\rm source}\rangle =0.79$. Adopting
$\sigma_v=993$\,km/s, the upper limit, leads to a critical curve at
$\sim 22\arcsec$. Hence we restrict our search to regions within a
radius of about $30\arcsec$, centred on the position of the
Bright Central Galaxy.\\
In a deep arc search using the WFPC2 archive, \cite{sand05a} quote
one arc in this archival HST data set (\tt A1 \rm in our data set).
Our identification of strong lensing features was done by visual
inspection of the WFPC2 frames in direct comparison with the deep
WFI exposures. As some of the candidates are very similar to not
fully removed cosmics we carefully searched for all identified
objects whether there is a corresponding object on all WFI frames.
In this way we identified 4 objects in total (denoted as \tt
A1,...,A4\rm) in the chosen field as good candidates for being
strong lensing features (see Figs.\,\ref{fig:HST_arcs}
and \ref{fig:arcdetail}).\\
A comparison even with shallow space based observations is a good
method to identify possible gravitational arcs due to the missing
atmospheric blurring effects. Several of the arc candidates were
smoothed on the WFI images so as to even lose their tangential
alignment. In particular, objects \tt A3 \rm and \tt A4 \rm are so
strongly influenced by observational effects that they are not
identifyable as arcs in ground based observations. A detailed
comparison between the WFI and WFPC2 images of the arcs is shown in
Fig.\,\ref{fig:arcdetail}. Note that the exposure time of the WFI
$R$-image is ~6.9\,h, whereas for the WFPC2 it was only
$\sim$0.28\,h. Nevertheless, the arc candidates visible in the HST
image are clearly recognizable as possible gravitational arcs,
whereas in the WFI frame seeing effects dominate the shape of the
objects.
%--------------------------------------------------------------------------
%--------------------------------------------------------------------------
%--------------------------------------------------------------------------
%--------------------------------------------------------------------------
\subsubsection{Determination of the length-to-width ratio}\label{subsubsec:l_to_w}
%--------------------------------------------------------------------------
To measure the length-to-width ratio $l/w$ we used \tt SExtractor
\rm to detect the arc candidates on the WFPC2 image as it is not
affected by atmospheric blurring. Due to its shallowness we used a
value of 0.75 for DETECT\_THRESH and ANALYSIS\_THRESH. The $l/w$
ratio itself was determined using the same ansatz as in
\cite{lenzen} and \cite{sextractor}: we treat the arcs as a set of
pixels with a certain light intensity value at each pixel. The light
distribution of a certain object is then defined by all
corresponding pixels detected by \tt SExtractor \rm shown in the
SEGMENTATION images. Hence we can compute the second moments
$\lambda_1$ and $\lambda_2$ of this light distribution in the usual
way \cite[see e.g.][]{lenzen,sextractor}. \rm Although the length
$l$ is not equal to $\lambda_1$ and the width $w$ is not equal to
$\lambda_2$ the \it ratio \rm $l/w$ is equal to
$\lambda_1/\lambda_2$ \citep{jaehne}. Hence we obtain the
length-to-width ratio by determining $\lambda_1$ and $\lambda_2$.
%--------------------------------------------------------------------------
\subsubsection{Photometry / catalogue creation}\label{subsubsec:arcphotometry}
\begin{figure}
   \centering
   \includegraphics[angle=0,width=9cm]{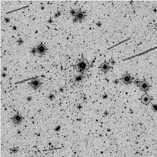}
      \caption{$R$-band image of Z3146. The image has a FoV of $\sim~16\arcmin40\arcsec\times~16\arcmin40\arcsec$
      (4.26$\;h_{70}^{-1}$\,Mpc$\times$4.26$\;h_{70}^{-1}$\,Mpc), the lines are the masks which were used to remove
      satellite tracks, foreground galaxies and bright halos from stars
       (see text for more details). North is up, East to the left.}
         \label{fig:864_masks}
\end{figure}

The photometry was also performed with the software package \tt
SExtractor 2.3.2\rm. In contrast to the determination of $l/w$ we
used the WFI frames for this purpose, as those images are much
deeper (see Table\,\ref{tab:data}). The photometric measurement on
the WFPC2 image was skipped as the F606W filter is fully
covered by the $V$ and $R$ band of the WFI observations. \\
As we concentrate on the cluster itself we restricted the extraction
of object catalogues to a FoV of
$\sim~16\arcmin40\arcsec\times~16\arcmin40\arcsec$
(4.26$\;h_{70}^{-1}$\,Mpc $\times$ 4.26$\;h_{70}^{-1}$\,Mpc in our
cosmology). The $V$ and $R$ images were convolved with a slight
Gaussian filter of width 0.61 and 0.91 pixels, respectively, to
bring all observations to the same seeing of $\sim1\myarcsec2$
($\Delta FWHM\leq10^{-3}$\arcsec) and hence ensure that all
objects are measured with the same photometric apertures. \\
We used \tt SExtractor \rm in double image mode with the deep $R$
band image as detection frame and the following parameters:
DETECT\_THRESH=7, ANALYSIS\_THRESH=7, and DETECT\_MINAREA=3 (the
higher detection threshold compared to the weak lensing analysis is
a result of the seeing correction). All magnitudes are obtained
using MAG\_AUTO with PHOT\_AUTOPARAMS=1,3.5, as elliptical apertures
and a Kron radius of this size is best suited to our observations.
In order to obtain clean catalogues with a minor fraction of
defective detections like obvious stars/foreground galaxies, traces
of asteroids and spurious detections in bright haloes of stars we
masked such objects to remove them from the final catalogues. The
image of Fig.\,\ref{fig:864_masks} shows as example the original
$R$-band image including all masked objects within a FoV of
$\sim~16\arcmin40\arcsec\times~16\arcmin40\arcsec$
(4.26$\;h_{70}^{-1}$\,Mpc $\times$ 4.26$\;h_{70}^{-1}$\,Mpc in our
cosmology). All masks were identical for the final $B$ and $V$
image, except for the individual satellite tracks. \\
The total galaxy catalogue contains 2138 objects having a
CLASS\_STAR parameter $<$ 0.95, MAG\_AUTO$<$99 in all bands
(considering an $E(B-V)=0.126$\,mag, taken from the
NED\footnote{http://nedwww.ipac.caltech.edu/}, based on
\cite{schlegel}) \rm and a FLUX\_RADIUS $>3.2$ pixels in $B$, $V$
and $R$, respectively. In addition we used WEIGHT maps created by
the data reduction pipeline \citep[see][for more
details]{pipeline1,pipeline2}. \\
%--------------------------------------------------------------------------
%_______________________________________________________________________________________________________________________________________
\subsubsection{Analysis of the strong lensing features}\label{strong_lensing}
\begin{figure*}
   \centering
   \includegraphics[angle=0,width=18cm]{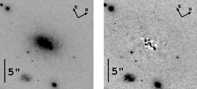}
      \caption{HST image of the innermost part of Z3146. In the right image we subtracted an elliptical
      model (based on fitting an ellipse to each isophote obtained by the \tt IRAF \rm tasks \it isophote \rm and
    \it bmodel \rm in the \tt STSDAS \rm package). Several
      distinct substructures are visible, but require deeper observations for their identification.}
         \label{fig:BCG}
\end{figure*}

\begin{figure}
   \centering
   \includegraphics[angle=0,width=9cm]{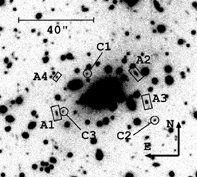}
      \caption{$R$-band image taken with WFI. Possible counter
      images of the arc candidates are denoted with \tt C\it x\rm. See
      Sect.\,\ref{strong_lensing} for the relations of the \tt C\it x \rm to the arc candidates.
        \label{fig:864R_counterarcs}}
\end{figure}

%--------------------------------------------------------------------------
The results of the photometric and morphological investigations of
all 4 arc candidates are summarised in Table\,\ref{tab:arcprops} and
Table\,\ref{tab:counterimages}, respectively. In this section we
analyse the arc candidates using these informations.\\
We can roughly estimate the strong lensing mass inside an Einstein
ring at the position of the outer most arc \tt A1 \rm
($\sim27\arcsec\sim120$\,kpc). This mass can be estimated to be
$M_{\rm sl}=1.39^{+0.36}_{-0.35}\times10^{14}M_\odot$, where the
main value is derived for $\langle z_{\rm source}\rangle =0.79$, the
mean redshift value obtained in Sect.\,\ref{sec:lensmass}. The
errors are calculated for $z_{\rm source}=[2\times z_{\rm
Z3146},2]$. However, as we do not know either the redshift or the
geometric alignment of the source with respect to the lens, this
procedure gives only a rough upper limit of the mass in the core.\\
One of the most interesting questions is the possibility of finding
multiple images of one single background source. Unfortunately we do
not have spectra of the objects (see
Sect.\,\ref{subsec:identifying}) which allow a secure identification
of counter images. Hence we search for
counter images in the following way:\\
Counter images of arcs may not appear as elongated objects in the
case of a folded arc system. In addition, they can differ in
magnitudes due to the gravitational magnifying effect and can appear
in unexpected locations \citep{broadhurst05a}, which are not
predictable without a precise model. Hence we have to restrict the
identification of multiple lensed objects to investigations of the
colour information ($B-V$), ($B-R$), and ($V-R$) only,
as they are conserved by lensing.\\
The search for multiple images was performed for all 4 arc
candidates independently in 2 steps: first, we searched the galaxy
catalogue for objects with (a) coinciding colours $(V-R)$, $(B-R)$
and $(B-V)$, and (b) lying in a radius of 30\myarcsec0 with respect
to the cluster center position in the RBS. In a second step we
discarded all objects being obvious cluster or
foreground galaxies by visual inspection. \\
\it arc candidate \tt A1\rm: In total we found two objects which
might be counter images of candidate \tt A1\rm, denoted by \tt
C1\rm, and \tt C2\rm, respectively (see
Fig.\,\ref{fig:864R_counterarcs}). It is hard to judge whether those
objects are counter images as they are hardly visible in the shallow
WFPC2 image. However, \tt C2 \rm is located at a distance of
$\sim30\arcsec$ with respect to the cluster center and hence we
expect it to be much more sheared at this position if it had
originated from the same object as \tt A1\rm. Additionally, the
colours agree only within their large error bars (see
Table\,\ref{tab:counterimages} for the numbers). Hence we conclude
that it is quite unlikely that
this object is a counter image of \tt A1\rm.\\
\it arc candidates \tt A2 \rm and \tt A3\rm: Both candidates show
colour coincidences with each other and \tt C1 \rm - \tt C3\rm.
However, again the colour differences only agree within their error
bars (see Table\,\ref{tab:counterimages} for the numbers). The
colour of \tt A2 \rm might be reddened to a certain amount by
elliptical galaxies in its vicinity,
nevertheless it is very unlikely that these objects have the same source.\\
\it arc candidate \tt A4\rm: We did not find any counter
image candidates for this object.\\[2mm]
Counter images also often occur nearby the central galaxy in the
case of a not perfect alignment between the observer, the lens and
the source. Hence they lie in the halo of a bright central object
affecting their colour and/or are dramatically sheared up to a
radial arc \citep[see][for some examples]{sand05a}.
Therefore we investigated the BCG using the HST image in more detail.\\
A closer look at the WFPC2 exposure of the BCG reveals some knots in
its very central part. To investigate these structures and to look
for a possible radial arc we subtracted as a first step an
elliptical model derived by fitting ellipses to isophotes of the BCG
(done with the help of the \tt IRAF \rm tasks \it isophote \rm and
\it bmodel \rm in the \tt STSDAS \rm package) as well as an
artificial de Vaucouleur profile
(task \it mkobjects \rm in \tt noao.artdata\rm).\\
Fig.\,\ref{fig:BCG} shows the central part of Z3146 with and without
the subtracted elliptical isophote model of the BCG. The removal of
the BCG reveals, apart from several clumps, an elongated
substructure in the centre of the BCG along its major axis in the
opposite direction to \tt A1\rm. However, we need deeper
observations for identification of this object. At the current stage
we can neither exclude the possibility of this object of being a
radial arc or a filamentary structure common in cooling flow clusters.\\

However, the fact that we did not find definitive counter images in
our observations does not mean that there are none. Lensed sources
can appear as very faint and thin arclets which are only visible in
deep HST observations. Such arclets are therefore hard to find in
ground based observations. Some prominent examples of lensing
clusters such a large number of faint arc(lets) are e.g. A2218
\citep[see][and references therein]{soucail04a}, A1689
\citep{broadhurst05a}, A370 \citep{bezecourt99a} or CL0024+16
\citep{broadhurst00a,kneib03a}.\\
%--------------------------------------------------------------------------
%--------------------------------------------------------------------------

%--------------------------------------------------------------------------

%--------------------------------------------------------------------------
\begin{table*}
\begin{center}
\begin{tabular}{c|c|c|c|c|c|c|c}
%\hline
arc candidate & angular distance  & projected dist.  & length $l$ & $l/w$ & \multicolumn{3}{c}{filter}\\
%\cline{6-11}
in Fig.\ref{fig:HST_arcs} & to cc  & to cc [kpc] & [ $\arcsec$ ] & WFPC2  & $B$ [mag] & $V$ [mag] & $R$ [mag]\\
\hline
~&~&~&~&~&~&~&~\\
\tt A1\rm & $\sim27$ & $\sim117$ & 2\myarcsec3 & $5.1$ & $24.49\pm0.28$ & $23.87\pm0.06$ & $22.73\pm0.01$\\[1mm]
\tt A2\rm & $\sim20$ &  $\sim87$ & 2\myarcsec0 & $5.6$ & $23.89\pm0.22$ & $23.07\pm0.04$ & $22.06\pm0.01$\\[1mm]
\tt A3\rm & $\sim23$ & $\sim100$ & 1\myarcsec4 & $2.1$ & $24.20\pm0.21$ & $23.68\pm0.05$ & $22.71\pm0.01$\\[1mm]
\tt A4\rm & $\sim26$ & $\sim113$ & 1\myarcsec1 & $1.9$ & $24.64\pm0.22$ & $24.08\pm0.05$ & $23.64\pm0.02$\\[1mm]
\hline
\end{tabular}
\end{center}
\caption{Table of the arc candidates shown in
Fig.\ref{fig:HST_arcs}. 'cc' denotes the cluster centre, the
length-to-width ratio $l/w$ is calculated with the help of
$\lambda_1$, and $\lambda_2$ (the second order moments of the light
distribution) and \rm is measured on the WFPC2 images as it is not
affected by atmospheric blurring. A value of $=0.75$ was chosen for
both, the DETECT\_THRESH and ANALYSIS\_THRESH due to the shallowness
of the image. The magnitudes are instrumental WFI magnitudes in the
Vega system. All objects are photometrised using \tt SExtractor's
\rm MAG\_AUTO.\label{tab:arcprops}}
\end{table*}

\begin{table*}
\begin{center}
\begin{tabular}{c|cc|cc|cc}
%\hline \hline
%  & &  &   &   &   &  \\
object & $(V-R)$ & $\Delta(V-R)$ & $(B-R)$ & $\Delta(B-R)$ & $(B-V)$ & $\Delta(B-V)$\\
\hline
~&~&~&~&~&~&\\
\tt A1 \rm &  1.14  &  0.07  &  1.76  &  0.29  &  0.62  &  0.34\\[1mm]
\tt A2 \rm &  1.01  &  0.05  &  1.83  &  0.23  &  0.82  &  0.26\\[1mm]
\tt A3 \rm &  0.97  &  0.06  &  1.50  &  0.22  &  0.52  &  0.26\\[1mm]
\tt A4 \rm &  0.44  &  0.07  &  1.00  &  0.24  &  0.55  &  0.27\\[1mm]
\tt C1 \rm  & 1.15  &  0.12  &  2.37  &  0.79  &  1.22  &  0.87\\[1mm]
\tt C2 \rm  & 1.02  &  0.1  &  2.26  &  0.74  &  1.24  &  0.81\\[1mm]
\tt C3 \rm  & 0.89  &  0.1  &  2.74  &  1.06  &  1.85  &  1.12\\[1mm]
\hline
\end{tabular}
\end{center}
\caption{Photometric properties of the arc candidates and their
possible counter images. See Sect.\,\ref{strong_lensing} for more
details and discussions.\label{tab:counterimages}}
\end{table*}
%_______________________________________________________________________________________________________________________________________
%_______________________________________________________________________________________________________________________________________
\section{Investigations of the cluster light distribution}\label{sec:furtherinvestigations}
%--------------------------------------------------------------------------
\begin{figure}
%   \centering
\vbox{
   \includegraphics[angle=-90,width=9cm]{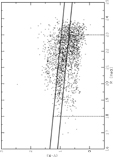}\\
   }
      \caption{$(V-R)$ vs. $R$ colour-magnitude diagram. A distinct Red Sequence is visible, framed by
      two solid lines. The dashed lines show the limits used for identification of cluster members (see text for
      more details).}
         \label{fig:3_colmag_plots}
\end{figure}
%--------------------------------------------------------------------------
%--------------------------------------------------------------------------
\begin{figure}
   \centering
   \includegraphics[angle=0,width=9cm]{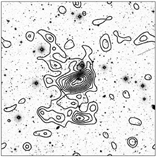}
      \caption{Galaxy density plot of Z3146. The FOV ($\sim16\arcmin40\arcsec\times16\arcmin40\arcsec$)
      is centred on the BCG, north is up, east to the left,
      $1\myarcsec0=4.36\;h_{70}^{-1}\,$kpc. The image shows the distribution
      of the main Red Sequence galaxies shown in the upper plot of
      Fig.\ref{fig:3_colmag_plots}.
      The contour levels correspond to
      $0.6, 0.8, 1, ..., 2.4\times10^{-4}$\,galaxies/arcsec$^2$.
      (The small dip in the contours between the cluster
    center and the bright star in the north is most likely an artifact
    from the masked and hence removed parts of the image, see
    Fig.\,\ref{fig:864_masks})        }
         \label{fig:galaxydensityplot}
\end{figure}
%--------------------------------------------------------------------------
In this section we present additional optical investigations on Z314
6 which are based on the WFI frames.
%--------------------------------------------------------------------------
\subsection{Cluster member catalogue}\label{subsubsec:catalogues}
%--------------------------------------------------------------------------
Independent of the previous analyses we have created different
catalogues as we have different selection criteria for the further
investigations. The lensing analysis focuses on background objects,
whereas the following investigations deal with the cluster members.
We extracted a catalogue of cluster members in the
following way:\\
(a) From the galaxy catalogue created in
Sect.\,\ref{subsubsec:arcphotometry} we made the colour-magnitude
diagram $(V-R)$ vs. $R$ (see Fig.\,\ref{fig:3_colmag_plots}). In
this plot we identify a Red Sequence (henceforth RS, marked by the
two solid lines) which is used as the basis for the cluster member
detection. The extraction of the Red Sequence was done by eye. As
the RS galaxies belong to the ellipticals, which are the reddest
ones in a galaxy cluster, we use the upper limit of the RS
distribution as the natural colour border and assume all objects
below the upper RS limit as cluster members. Additionally we skipped
all objects with $R\geq23.5$\,mag as likely belonging to a
background population, and objects with $R\leq18$\,mag as likely
foreground systems. With these criteria we found in total 756 RS
galaxies and 1478 cluster members.

%--------------------------------------------------------------------------
\subsection{Galaxy distribution in Z3146}\label{subsubsec:morphology}
%--------------------------------------------------------------------------

To investigate the distribution of the RS members we created galaxy
density maps in the following way: a blank image of about
$4200\times4200$ pixels (corresponding to a FoV of
$\sim~16\arcmin40\arcsec\times~16\arcmin40\arcsec$,
4.26$\;h_{70}^{-1}$\,Mpc $\times$ 4.26$\;h_{70}^{-1}$\,Mpc) was
created with pixel value "0" everywhere. At each position of the
extracted Red Sequence galaxies (see Sect.\,\ref{sec:opt_analysis})
the pixel value was changed to "1" and a subsequent Gaussian
smoothing with $\sigma=241$ pixels (corresponds to
250$\;h_{70}^{-1}$\,kpc) leads to the image in
Fig.\ref{fig:galaxydensityplot}.\\
The galaxy density plot of the main Red Sequence
(Fig.\ref{fig:galaxydensityplot}) shows one large peak centred on
the main cluster with no distinct subclumps, except one small peak
south of the cluster core. This is an indication that Z3146 is a
relaxed cluster without any ongoing major merger event, which is
confirmed by the massive cooling flow found in previous
investigations \citep{edge94a,fabian02a} and our own results of
$\sim 1600$\,M$_\odot$ per year. In particular, the small distance
of about $0\myarcsec69$ between the optical and the
X-ray centre \citep{schwope} also confirms the calm character of this cluster.\\
%_______________________________________________________________________________________________________________________________________
\subsection{Light distribution / mass-to-light ratio}\label{subsubsec:masstolight}
%_______________________________________________________________________________________________________________________________________
In order to obtain a mass-to-light ratio and creating a light
distribution map we applied the K-correction as a first step. We
used the MatLab$^\copyright$ script \tt lum\_func.m \rm written by
Eran Ofek\footnote{\tt\scriptsize
http://wise-obs.tau.ac.il/$\sim$eran/matlab.html} for this purpose.
As input parameters we used the corresponding WFI filter
curves\footnote{\tt\scriptsize
http://www.ls.eso.org/lasilla/sciops/2p2/E2p2M/WFI/filters/} and
template spectra provided by Stephen Gwyn\footnote{\tt\scriptsize
http://orca.phys.uvic.ca/$\sim$gwyn/pz/specc/} which are based on
spectra by \cite{coleman80a}. As Red Sequence galaxies are mainly
ellipticals we used E/S0 spectra for them and Sbc templates for the
remaining. The resulting K-corrections (see
Table\,\ref{tab:k-correction}) were applied to the galaxy catalogues
of all galaxies in the FoV. To take the contamination resulting from
non-cluster members into account we created another catalogue of
galaxies with the same criteria from a different region on the final
WFI frames. This field is centered on $\alpha=10^{\rm h}22^{\rm
m}59^{\rm s}$ and $\delta=+04^\circ20\arcmin33.3\arcsec$, has the
same size as the region used for creating the cluster member
catalogue and has no distinct galaxy density peak. Hence we assume
the galaxy population in this field to be dominated by field
galaxies. Additionally this region is, in spite of the pointing
offset between the different observing programs (see
Sect.\,\ref{sec:opt_observations}), visible on all three observed
bands. The galaxy counts in this field were binned in the same way
as in the cluster field and subtracted from the
corresponding bin of the cluster count.\\
\begin{table*}
\begin{center}
\begin{tabular}{c|c|c|c|c}
& $B$ & $V$ & $R$ & units\\
\hline
%\cline{6-11}
~&~&~&~&~\\
E/S0 K-correction       & 1.41 & 0.85 & 0.32 & [mag]\\
Sbc K-correction        & 0.84 & 0.33 & 0.14  & [mag]\\
fitting range $M_{fit}$ & $<-19$ & $<-19$ & $<-19$ & [mag]\\[1mm]
$\phi^*$                & $1.814^{+2.24}_{-2.24}\times10^3$ & $0.8^{+2.81}_{-2.81}\times10^3$  & $0.907^{+1.121}_{-1.121}\times10^3$  & [\# galaxies deg$^{-2}$]\\[1mm]%^{}_{}
$\alpha$                & $-1.136^{+0.53}_{-0.53}$ & $-1.38^{+1.30}_{-1.30}$ & $-1.24^{+0.33}_{-0.33}$  \\[1mm]
$M^*$                   & $-20.6^{+1.3}_{-1.3}$ & $-21.74^{+3.01}_{-3.01}$ & $-22.53^{+1.55}_{-1.55}$   & [mag]\\[1mm]
$\chi^2$                & 0.972 & 0.952 & 0.973 \\[1mm]
$L_{tot}$               & $1.12\times10^{14}$ & $0.98\times10^{14}$ & $1.38\times10^{14}$  & $L_\odot$\\
$F$                     & $<1\%$ & $<1\%$ & $<1\%$ & \\
\end{tabular}
\end{center}
\caption{Details of the catalogues, K-corrections values and the
Schechter function parameters $\phi^*$, $\alpha$ and $M^*$ for the
three WFI filters. $F$ is the fraction of light which is missing due
to the limiting magnitudes $M_{\rm lim\it}$ of the catalogue given
in Table\,\ref{tab:data}. For the $R$ band we used the limiting
magnitude for region A (see Table\,\ref{tab:data}). See text for
more details. The Luminosity distance is 1.5Gpc (distance modulus
$m-M=40.88$ at $z=0.2906$). \label{tab:k-correction}}
\end{table*}
%--------------------------------------------------------------------------
%%--------------------------------------------------------------------------
%\begin{figure}
%%   \centering
%\vbox{
%   \includegraphics[angle=-90,width=9cm]{lowres/B_LumFunc.ps}\\
%   \includegraphics[angle=-90,width=9cm]{lowres/V_LumFunc.ps}\\
%   \includegraphics[angle=-90,width=9cm]{lowres/R_LumFunc.ps}\\
%}
%      \caption{Luminosity functions of $B$, $V$ and $R$. The lines are fitted Schechter functions
%      (see text for more details).\bf xxxx THIS FIGURE WILL BE REMOVED, BUT KEPT BY THE MOMENT TO KEEP THE NUMBERING OF FIGURES xxxx \rm}
%         \label{fig:mag_distr}
%\end{figure}
%--------------------------------------------------------------------------
%--------------------------------------------------------------------------
\begin{figure}
   \centering
   \includegraphics[angle=0,width=9cm]{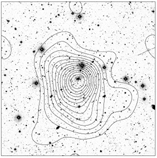}
      \caption{$V$ band light distribution map. The FoV is
      $\sim16\arcmin40\arcsec\times16\arcmin40\arcsec$, North is
      up, East to the Left. The contours correspond to values
      $2\times10^{5}$, $2.5\times10^{5}$, $3\times10^{5}$,
      ..., $7.5\times10^{5}$, $8\times10^{5}$\,L$_\odot$.
      The light distribution shows one
      single peak without any distinct
      substructures. The position difference between the cluster
    center and the peak of the distribution is about $\sim 29\arcsec$,
        which might be a result from the masked parts of the image, see
Fig.\,\ref{fig:864_masks})}
         \label{fig:light_distr}
\end{figure}
%--------------------------------------------------------------------------

%%%%%%%%%%%%%%%%%%%%%%%%%%%%%%%%%%%%%%%%%%%%%%%%%%%%%%%%%%%%%%%%%%%%%%%%%%%%%%
To calculate the total luminosity in the $V$ band in solar units we
assume the solar absolute magnitude to be $\rm M\it_V=$4.82mag
\citep{cox}. We also assume the cluster members to follow the
standard Schechter luminosity function \citep{schechter76a}:
\begin{equation}\label{eq.schechter}
    \phi(L)dL=\phi^*\left(\frac{L}{L^*}\right)^{\alpha}e^{-\left(\frac{L}{L^*}\right)}d\left(L/L^*\right)
\end{equation}
In terms of magnitudes the Schechter function reads
\begin{equation}\label{eq.schechter_mag}
    \phi(M)dM=0.4\ln(10)\cdot\phi^*\cdot10^{-0.4(M-M^*)(\alpha+1)}\cdot
    e^{\left[-10^{-0.4(M-M^*)}\right]}dM
\end{equation}
To obtain the parameters $L^*$, $\phi^*$ and $\alpha$ we applied a
$\chi^2$ fit to Eq.\ref{eq.schechter_mag} using the
MatLab$^\copyright$ Fitting Toolbox at a 95\% confidence level. The
best fitting parameters including the $\chi^2$-value for the
goodness of the fit and the fitting range for the luminosity
$[\inf,L_{fit}]$ are given in Table\,\ref{tab:k-correction}, where
$L_{fit}$ is the completeness luminosity.\\
%Fig.\,\ref{fig:mag_distr} shows the fitted Schechter functions: The
%dashed lines are the best fitting
%curves obtained with the parameters $L^*$, $\phi^*$ and $\alpha$ given in Table\,\ref{tab:k-correction}.\\
The total luminosity $L_{tot}$ (see Table\,\ref{tab:k-correction})
can now be obtained by integrating Eq.\,\ref{eq.schechter}:
\begin{equation}\label{eq.integratedschechter}
    L_{tot}=\phi^*L^*\Gamma(2-\alpha)
\end{equation}
The integration of the Schechter function down to a luminosity
$L_{\rm lim\it}$ is equal to
\begin{equation}
    L_{lim}=\phi^*L^*\Gamma(2-\alpha,L_{lim}/L^*).
\end{equation}
As the Schechter function is not applicable to very faint
luminosities we calculate the fraction of light we miss due to
observational effects
\begin{equation}
    F=1-\frac{\Gamma(2-\alpha,L_{fit}/L^*)}{\Gamma(2-\alpha,L_{lim}/L^*)}
\end{equation}
$L_{fit}$ being the luminosity (corresponding to $M_{fit}$ in
Table\,\ref{tab:k-correction}) down to which our catalogues are
complete and $L_{lim}$ is the limiting luminosity corresponding to
the limiting magnitude given in Table\,\ref{tab:data}.\\
Fig.\ref{fig:light_distr} shows the corresponding light distribution
of the Red Sequence galaxies in $V$. Again, only one single, very
distinct peak centred on the BCG is visible. In addition the
distribution is very smooth and shows no substructures, confirming
the relaxed
state of Z3146.\\[0.5cm]
In Sect.\,\ref{sec:lensmass} we fitted the mass obtained by the weak
lensing method to an NFW profile with the following best-fit
parameters obtaining $r_{200}=1661^{+280}_{-328}\;h_{70}^{-1}$kpc
and $c=3.6^{+2.8}_{-2.4}$ as best fit values. This leads to
$M_{200}=7.04^{+6.03}_{-2.32}\times10^{14}M_{\odot}$. Using these
values we find a mass-to-light ratio within $r_{200}$ in the $V$
band of $M/L_V\sim156^{+134}_{-51}M_\odot/L_\odot$. This value is in
agreement with e.g. \cite{hradecky00a}, who give a median value of
$M/L_V\sim138M_\odot/L_\odot$ for eight clusters within a radius of
$\sim1.38\,h_{70}^{-1}$Mpc.

%______________________________________________________________
%\section{Optical vs. X-ray}
%______________________________________________________________

\section{Discussion and summary}\label{sec:summary}
We presented a combined investigation of optical and X-ray
observations of the prominent galaxy cluster Z3146. This cluster
seems to be in a relaxed state, which is confirmed by
\begin{itemize}
\item the absence of large substructures in the galaxy density plot
and the light distribution map (see
Fig.\,\ref{fig:galaxydensityplot} and Fig.\,\ref{fig:light_distr}),
\item the regular shape of the
cluster in the X-ray image and the temperature map (see
Fig.\,\ref{image} and \ref{tmap.fig}, respectively)
\item the massive nominal
cooling flow of $\sim1600M_\odot$yr$^{-1}$
\item the good coincidence
of the optical, the X-ray, and the weak lensing centre (each of the
order of a few arcseconds), and
\item the regular shape of the weak lensing
mass reconstruction.
\end{itemize}
Further optical investigations on the cluster also revealed four
gravitational arc candidates and a mass-to-light ratio of
$M/L_V\sim156^{+134}_{-51}\,\,h_{70}^{-1}M_\odot/L_\odot$ at
$r_{200}=1661^{+280}_{-328}\;h_{70}^{-1}$kpc.\\

\begin{figure}
   \centering
   \centerline{\includegraphics[angle=0,width=\hsize]{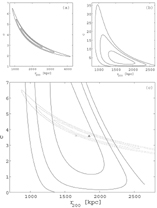}}
   \caption{Comparison of confidence levels from NFW fits: Panel (a) shows the fit to the X-ray data,
    Panel (b) the fit to the shear data. A direct comparison with the corresponding best fit values is
    presented in Panel (c), where the lensing model is represented by the solid lines (best fit marked by 'x'),
    and the X-ray model by the dash-dotted
    lines (best fit values correspond to the triangle). See Sect.\,\ref{sec:summary} for more details.}
         \label{fig:nfwcontours}
\end{figure}

%_______________________________________________________________________________________________________________________________________
%\subsubsection{Comparison of the X-ray- and Lensing based NFW Models}\label{subsec:nfwcomparison}
%--------------------------------------------
We also determined the mass of this cluster with two independent
methods, weak lensing and X-ray measurements. Both data sets, X-ray
and lensing, were used to establish best fits to the commonly used
NFW model. Fig.\,\ref{fig:nfwcontours}
shows the comparison of the confidence levels for these data:\\
{\it Fig.\,\ref{fig:nfwcontours}(a) - Xray-data}: Confidence level
from a NFW fit to the mass profile derived from X-ray data according
to Eq. \ref{massbeta.eq}. The contour levels are 2.31, 6.25, 11.90
corresponding to confidence levels of 68.3\% ($1\sigma$), 95.4\%
($2\sigma$), 99.73\% ($3\sigma$).\\
{\it Fig.\,\ref{fig:nfwcontours}(b) - Lensing data}: The contours
are at $2\Delta{\cal L}=$2.30, 6.17, 9.21 corresponding to
confidence levels of 63.8\%, 90\%, 95.4\% and 99\% if we assume
Gaussian statistics. We varied the galaxy sample of the lensing fit
to investigate the dependence of the result on this parameter. On
the one hand, lowering the maximum radius to which galaxies enter
the calculations to $8\myarcmin 0$ from the galaxy centre (we have
full data coverage around the cluster up to this radius; see Fig.
\ref{fig:layout}) leads to $r_{200}=1610\;h_{70}^{-1}$kpc; $c=4.01$.
On the other hand, not considering the inner parts of the cluster
and using only galaxies with a distance larger than $2\myarcmin 0$
we obtain $r_{200}=1748\;h_{70}^{-1}$kpc; $c=1.0$. We note that
$r_{200}$ is reasonably well constrained and that the concentration
$c$ mainly depends on the details near the cluster core.  This
behaviour corresponds to the shape of our contours and is typical
for NFW profile fits in weak lensing studies; see e. g.
\cite{clowe02a} and \cite{dietrich05a}.  The parameter ranges in $c$
and $r_{200}$ imply an uncertainty of the total cluster mass of
$10\%-20\%$ (considering radii of $1-2\;h_{70}^{-1}$Mpc; see also
Fig.\,\ref{fig:lensmass}).\\
{\it Fig.\,\ref{fig:nfwcontours}(c) - Direct Comparison}: The 'x'
marks our best fit lensing value of $r_{200}=1661\;h_{70}^{-1}$\,kpc
($r_{s}=460.1\;h_{70}^{-1}$\,kpc) and $c=3.61$, which lies in the
vicinity of the $3\sigma$ X-ray model. The triangle corresponds the
best NFW fit to the X-Ray data: $r_{200}=1849\;h_{70}^{-1}$\,kpc,
$c=3.58$ ($r_s=516.6\;h_{70}^{-1}$\,kpc, see Sect.\,\ref{mass.sec}
for more details). This best fit value is located within the
$1\sigma$ contour of the lensing model. Hence both models are in
excellent agreement.

%--------------------------------------------------------------------------

\begin{figure}[ht]
   \centering
   \includegraphics[angle=0,width=9cm]{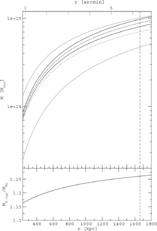}
      \caption{Comparison of the best fit NFW models obtained from X-ray data
      (black lines) and weak lensing signal (gray lines). Both show good
      coincidences within 1$\sigma$ errors (see Fig.\,\ref{fig:nfwcontours}). In the lower panel the
      ratio of $M_{X-ray}/M_{wl}$ is shown. Note, that the fit of the
      X-ray data to a NFW profile starts at $\leq230\;h_{70}^{-1}$kpc.
      The dash-dotted line represents $r_{200}=1661h_{70}^{-1}$kpc.
      See Sect.\,\ref{sec:summary} for more details.  }
         \label{fig:masscomparisons}
\end{figure}

A direct comparison of the mass profiles and the ratio between
$M_{X-ray}/M_{wl}$ is given in Fig.\,\ref{fig:masscomparisons},
which shows that the best fit models agree within $\sim20\%$.\\
Comparing the strong and weak lensing masses it seems that they
disagree. At the radius of the outermost arc at $r\sim27\arcsec$
($\sim120$\,kpc) we obtain a mass within an Einstein ring of about
$M_{\rm
sl}=1.39^{+0.36}_{-0.35}\times10^{14}\,\,h_{70}^{-1}$\,M$_\odot$ for
the strong lensing measurement. The NFW profile of the weak lensing
fit gives $M_{\rm
wl}=2.35^{+2.58}_{-1.56}\times10^{13}\,\,h_{70}^{-1}$\,M$_\odot$ at
the same position, which is, the best case assuming, roughly half of
$M_{\rm sl}$ only. However, due to the large uncertainties in the
strong lensing mass determination (unknown redshifts, unknown
lensing geometry...) we assume this mass only to be a rough upper
value. Hence this discrepancy is likely an artifact of the large
numbers of uncertainties in the determination of the strong lensing
mass.
%--------------------------------------------------------------------------
%--------------------------------------------------------------------------

Especially in relaxed clusters, the mass estimates obtained from
weak lensing and X-ray mass methods usually seem to agree very well
\citep{allen98a,wu98a}. Recent observations of cooling flow clusters
derived from \it Chandra \rm and/or \it XMM-Newton \rm confirm these
results \citep[see e.g.][]{allen02a,cypriano05a}. We find a
temperature of $5.9\pm0.1$\,keV in Z3146, in agreement with the
assumption of \cite{cypriano04a} that clusters having an ICM
temperature $\lesssim8.0$\,keV are in a relaxed state. In particular
relaxed clusters are interesting for cosmological studies as their
mass content tends to take a spherically symmetric shape, which is
the usual assumption in theoretical approaches. Hence a large sample
of such systems is a useful probe to verify whether the mass density
of galaxy clusters follows an NFW profile \citep{nfw1,nfw2}, or
whether a different profile like the Burkert \citep{burkert00a}, the
Moore \citep{moore99a} or the non-extensive profile
\citep{leubner05a,kronberger06a}
is a suitable description.\rm\\

\begin{acknowledgements}
      We are very grateful to Ludovic van Waerbeke for his help
      with the weak lensing cluster mass reconstruction and thank
      Joachim Wambsganss and Peter Schneider for fruitful
      comments. The authors also want to thank Rocco Piffaretti
      for his kind help during the NFW fit of the X-ray data,
      Eran Ofek for providing very useful
      MatLab\copyright scripts and Leo Girardi for kindly
      generating isochrones for the WFI filters. We also thank the anonymous referee
      for invaluable comments, and S. Ettori
      for providing the software required to produce the X-ray
      colour map in Fig.\,\ref{tmap.fig}. This work is supported by
      the Austrian Science Foundation
      (FWF) project number 15868, by the Deutsche
      Forschungsgemeinschaft (DFG) under the project ER 327/2-1, by NASA grant
      NNG056K87G
      and by NASA Long Term Space Astrophysics Grant NAG4-11025.
\end{acknowledgements}

\bibliographystyle{aa}
\bibliography{Reflist}{}

\end{document}